%% file: main.tex
\documentclass{jpp}
\usepackage{graphicx}
\usepackage{xcolor}
\usepackage[utf8]{inputenc}
\usepackage[T1]{fontenc}
\usepackage{amsmath,breqn}
\usepackage{aas_macros}
\usepackage[colorlinks=true,citecolor=blue,urlcolor=blue,linkcolor=blue]{hyperref}
\usepackage{adjustbox}

\newcommand{\feq}[1]{f_{\mathrm{eq}}^{#1}}

\makeatletter
\let\cat@comma@active\@empty
\makeatother
\allowdisplaybreaks
\title{Kinetic magnetohydrodynamics and Landau fluid closure in relativity}

\author{Abhishek Hegade K. R. 
\aff{1}
\corresp{\email{ah4278@princeton.edu}}
\and 
James M. Stone\aff{2}
}

\affiliation{\aff{1}Princeton Gravity Initiative, Princeton University, Princeton, NJ 08544, USA
\aff{2}School of Natural Sciences, Institute for Advanced Study, Princeton, NJ, USA}

\begin{document}

\maketitle

\begin{abstract}
Diffuse accretion flows near a supermassive black hole are fundamentally weakly collisional. In such weakly collisional plasmas, the ion and electron distribution functions can deviate significantly from thermal equilibrium, and particle kinetic effects can influence large-scale fluid motion by driving pressure anisotropy, heat conduction, and plasma instabilities. Modeling these plasma effects in highly relativistic flows could be important for interpreting horizon-scale observations of black hole images. In this paper, we present a theoretical framework for understanding weakly collisional plasmas in general relativity by deriving the relativistic drift kinetic equations from the Vlasov-Maxwell equations. We present the evolution equations for the moments of the gyroaveraged distribution function and introduce a new analytic Landau fluid closure to capture anisotropic heat flow in relativistic plasmas. Unlike standard (collisional) general relativistic magnetohydrodynamics or extended magnetohydrodynamics, our model does not rely on strong collisions to enforce thermal equilibrium and consistently incorporates Landau damping in a fluid closure. The model introduced in this work provides a complementary approach to fully kinetic simulations in understanding weakly collisional effects in low-luminosity relativistic black hole accretion disks.
\end{abstract}
\section{Introduction}
Accretion around supermassive black holes (SMBHs) is a fundamental process that shapes galactic evolution and the growth of cosmic structures. Modeling the complex turbulent gas flow near the horizon of a SMBH remains a key problem in high-energy astrophysics~\citep{2014ARA&A..52..529Y,Blaes_2013,Abramowicz_2013}. This problem has recently gained unprecedented attention following the pioneering horizon-scale observations of the SMBHs in M87 and Sgr A* by the Event Horizon Telescope (EHT)~\citep{EventHorizonTelescope:2019dse,EventHorizonTelescope:2022wkp}. Interpreting EHT images typically relies on ideal general relativistic magnetohydrodynamics (GRMHD) to model the highly magnetized and relativistic gas flow near the event horizon~\citep{EventHorizonTelescope:2019pcy,Narayan:2021qfw,EventHorizonTelescope:2022urf,Dhruv:2024igk}. However, the extreme temperatures and sub-Eddington accretion rates characteristic of these sources suggest the plasma is fundamentally weakly collisional or collisionless~\citep{1997ApJ...490..605M,Chandra_2015}. In these regimes, the ideal fluid approximation breaks down as the mean free path for collisions exceeds the characteristic gravitational scale of the system, leading to significant non-ideal effects such as pressure anisotropy, heat conduction along magnetic field lines, and thermal disequilibrium in the ion and electron distribution functions~\citep{Galishnikova_2023,Vos-2025}.

When flow speeds are non-relativistic and gravity is Newtonian, there are well-developed models, such as Navier-Stokes theory and kinetic magnetohydrodynamics (KMHD), which can be used to model non-ideal effects~\citep{Landau:1987,2004ppa..book.....K}. The same situation is not true in relativity. Historically, the first approaches, such as those of Eckart \citep{Eckart:1940te} and Landau and Lifshitz \citep{Landau:1987}, generalized the non-relativistic Navier-Stokes theory to include heat conduction, bulk viscosity, and shear viscosity. It was realized that naively extending the non-relativistic Navier-Stokes theory to relativity leads to acausal behavior—due to the parabolic nature of the equations—and generic instabilities \citep{Hiscock:1985zz}. However, the recently developed Bemfica-Disconzi-Noronha-Kovtun (BDNK) theory \citep{Bemfica_2018, Kovtun2019, Bemfica2022} has demonstrated that these pathologies are not intrinsic to first-order derivative expansions but are instead a consequence of the ill-defined choice of variables, such as velocity and temperature out of equilibrium—the so-called choice of a ``hydrodynamic frame''. By employing a more general hydrodynamic frame than the Eckart and Landau-Lifshitz frames, BDNK showed that one could ensure causality and stability within a first-order framework.

Conversely, a more traditional route for maintaining finite propagation speeds in relativistic dissipation theories is to use a relativistic extension of Grad’s 14-moment approximation \citep{Grad:1949zza} to obtain the Mueller-Israel-Stewart (MIS) theories \citep{Muller-book, Israel:1979wp, Denicol_2012, Denicol_2018}. As a ``second-order'' theory, MIS promotes dissipative fluxes—such as the shear stress tensor and heat flux—to independent dynamical variables that evolve toward their Navier-Stokes values over finite relaxation timescales. In the black hole accretion context, this MIS framework has been adapted into the ``Extended MHD'' (EMHD) model by \cite{Chandra_2015, Foucart_2015, Foucart_2017, dhruv2025electromagneticobservablesweaklycollisional}. EMHD utilizes an MIS-type to incorporate non-ideal effects such as pressure anisotropy and thermal conduction along magnetic field lines, successfully recovers the collisional Braginskii theory \citep{1958JETP....6..358B} in the non-relativistic limit, is causal and stable under certain conditions \citep{Cordeiro:2023ljz} and has been extended to include multi-species flows~\citep{Most_2021, Most_2022}.

The dissipative theories discussed above, whether based on the first-order BDNK or second-order MIS/EMHD framework, rely on collisions to mediate non-ideal effects and require all plasma species to be close to local thermal equilibrium. 
This assumption can fail for sources like M87 or Sgr A*~\citep{2009ApJ...699..626H}, where a more direct link to collisionless particle kinetics is needed to capture strong deviations from thermal equilibrium and to model kinetic instabilities such as the mirror and fire-hose instabilities~\citep{1993tspm.book.....G,PhysRevLett.100.065004,Kunz_2014}.
The first-principles approach to modeling kinetic effects is to numerically evolve the distribution function of each species in the plasma using particle-in-cell (PIC) simulations. While such PIC simulations provide a microscopic picture of kinetic effects, the vast separation of scales between the particle gyroradius and macroscopic fluid motion makes these calculations highly computationally intensive~\citep{Galishnikova_2023, Vos-2025}.

\begin{table*}
    \centering
\begin{tabular}{ c  c  }
    \hline
        \textbf{Description} & \textbf{Equation number } \\
        \hline 
        Drift kinetic equation & Equation~\eqref{eq:gyroaveraged-kinetic-equation}.
         See Sec.~\ref{sec:KMHD-summary}  \\
         &for a summary of the drift kinetic equations. \\
        \hline
        Evolution of the gyroaveraged moments 
        & Equations~\eqref{eq:moment-equations-general} and \eqref{eq:all-species-equations-3+1}. \\
        \hline 
        Relativistic plasma response function & Equation~\eqref{eq:Z-definition} and Appendix~\ref{appendix:plasma-response-function}.
        \\
        \hline 
        KMHD linear response & Equation~\eqref{eq:linearized-moments-analytical}.
        \\
        \hline 
        Closure for heat fluxes & Equation~\eqref{eq:closure-analytical-heat-fluxes}.
        \\
        \hline 
\end{tabular}
\caption{List of the most important equations in the paper.}
    \label{tab:list-of-imp-eqns}
\end{table*}
For non-relativistic fluid flows, a common macroscopic approach to model weakly collisional plasmas is to use the KMHD equations~\citep{1983bpp..conf....1K}. In this approach, a strong magnetic field provides effective collisionality and cohesion, forcing the kinetic plasma to behave like a fluid. The strong magnetic force confines the particles to gyrate along the magnetic field lines. Assuming a large-scale separation between the particle gyroradius and the macroscopic fluid motion allows one to obtain an evolution equation for the gyroaveraged plasma distribution function. Combining conservation laws for number density, momentum, and energy with quasi-neutrality yields a set of kinetic equations coupled to hydrodynamic equations that can be evolved to understand the effect of particle kinetics on the fluid flow.
In practice, solving the gyroaveraged kinetic equation is almost as expensive as a PIC simulation. 
Moreover, determining the parallel electric field and ensuring that the firehose and mirror instabilities evolve in a consistent manner requires careful numerical treatment, see Sec. 5 of~\citet{Juno_2025}. 
Therefore, one usually uses the moments of the gyrokinetic equation understand the evolution of thermodynamic variables, such as the pressure parallel and perpendicular to the magnetic fields.
A consistent formulation of the moment equations for the pressure variables requires one to obtain closure relations for the heat fluxes, which can reproduce the behavior of the kinetic equation and capture the Landau damping in the system~\citep{PhysRevLett.64.3019,1997PhPl....4.3974S,Goswami_2005,Hunana_2019}.

The main goal of this paper is to extend the KMHD framework to general relativity. We do this by analyzing the Vlasov-Maxwell equations in a general curved spacetime and make no approximation on the nature of the distribution function or the stress energy tensor of the fluid~\citep{1991PhFlB...3.1871G,Gedalin-1995}.
Assuming a large scale separation between particle kinetics and macroscopic fluid motion allows us to derive the drift kinetic equations for the gyrotropic distribution function directly from the Vlasov-Maxwell equations.
We also analyze the linear plasma response of relativistic plasma in detail and provide definitions for the relativistic plasma response function.
By matching the linear kinetic response with a fluid response~\citep{1997PhPl....4.3974S} we provide a consistent closure relationship for the heat fluxes parallel and perpendicular to the magnetic field.
Our closure relationship is analytic in the ultra-relativistic limit, and does not rely on a Chapman-Enskog-like collisional expansion of the distribution function~\citep{2008PhPl...15f2112T}.

The theoretical framework derived here should complement the PIC approach in understanding the gyrokinetic effects in black hole accretion and provide a route to understand the effects of pressure anisotropy~\citep{Sharma_2003,Sharma_2006,Kunz_2014}, turbulence~\citep{2009ApJS..182..310S,Kunz_2016}, and magnetic immutability~\citep{Squire2023} in a relativistic context.
The rest of the paper describes our results in detail and is organized as follows. In Sec.~\ref{sec:KMHD-section-main}, we present the derivation of the KMHD equations in general relativity, starting from the Vlasov-Maxwell equations. We also discuss the evolution equations for the moments of the distribution function and the Chew-Goldberger-Low (CGL) equations~\citep{CGL-Original} in relativity.
Section~\ref{sec:linear-response} analyzes the linear response of an initially anisotropic plasma and provides analytic expressions for the evolution of the moments of the distribution function.
We then present a Landau fluid closure in the ultra-relativistic limit in Sec.~\ref{sec:Landau-Closure} and show that our closure captures the Landau damping.
We present our conclusions in Sec.~\ref{sec:conclusions} and relegate most of the technical details to the appendix.
We use geometric units throughout the paper and use the mostly positive metric signature.
A list of the most important equations in the paper is presented in Table~\ref{tab:list-of-imp-eqns}.
\section{Kinetic MHD equations in general relativity}\label{sec:KMHD-section-main}
In this section, we derive the KMHD equations in relativity starting from the Vlasov-Maxwell equations.
We set up our notation and recall the Vlasov-Maxwell equations in general relativity in Sec.~\ref{sec:Vlasov-Maxwell-Basics}.
Next, we perform a formal expansion in gyro-radius to obtain the kinetic-MHD equations in Sec.~\ref{sec:KMHD-section}.
In Sec.~\ref{sec:moment-equations}, we derive the evolution equations for the moments of the distribution function, focusing on the evolution equations for the pressure parallel and perpendicular to the magnetic field.
We discuss the double adiabatic limit of the evolution equations in Sec.~\ref{sec:double-adiabatic-limit}.
Readers interested in the final equations can skip directly to Sec.~\ref{sec:KMHD-summary}, where we summarize the important equations obtained in this section.
A short summary of different hydrodynamic frames and the relationship between MHD and force-free dynamics is provided in Appendix~\ref{appendix:hydro-frames}.

Before we begin, we set up some notation.
In this section, we study the Vlasov-Maxwell equations on a fixed four-dimensional spacetime with metric $g$. We assume the plasma has $S$ different particles.
The charge, mass, and one-particle distribution function of species $s$ are denoted by $q_{s}$, $m_{s} >0$, and $f_{s}$, respectively. We denote the electromagnetic field tensor by $F_{\alpha \beta}$. Instead of working on the phase space $(x,p)$ where $p^{\mu}$ is the momentum, we work in $(x,k)$ space where $k^{\mu} = p^{\mu}/m_s$ is the velocity of the particle.
Given a function $\mathcal{I}(x,k)$, we use $\left< \mathcal{I} \right>_{s}$ to denote its average with respect to the distribution function $f_s$
\begin{align}\label{eq:bracket-s-def}
    \left<\mathcal{I}\right>_{s} \equiv \int d^4 k \sqrt{-g}  f_{s} \mathcal{I}(x,k) \,.
\end{align}
We also sometimes use $\left< \mathcal{I}\right>$ without any subscript to denote the average of the function over the velocity space
\begin{align}\label{eq:bracket-without-s-def}
    \left<\mathcal{I}\right> \equiv \int d^4 k \sqrt{-g} \mathcal{I}(x,k)
    \,.
\end{align}
\subsection{Vlasov-Maxwell equations}\label{sec:Vlasov-Maxwell-Basics}
The Vlasov-Maxwell equations on phase space $(x,k)$ are given by~\citep{Sarbach2013}
\begin{subequations}
\begin{align}
    \label{eq:Vlasov-Maxwell-Eqn}
    &k^{\mu} \frac{\partial f_{s}}{\partial x^{\mu}}
    +
    \left[- \Gamma^{\alpha}_{\mu \nu} k^{\mu} k^{\nu} + \frac{q_{s}}{m_{s}} F^{\alpha}{}_{\mu} k^{\mu} \right]
    \frac{\partial f_{s}}{\partial k^{\alpha}}
    =
    \mathcal{C}_s\left[f_s\right]
    \,,\\
    \label{eq:Gauss-Farday-laws}
    &\nabla_{\mu} (\star F^{\mu \nu}) = 0\,, \quad 
     \star F^{ \mu \nu} = \frac{1}{2} \epsilon^{\mu\nu\kappa\lambda} F_{\kappa \lambda}\,,\\
    \label{eq:EM-equations}
   & \nabla^{\mu} F_{\mu \nu}
   = -4 \pi \sum_{s=1}^{S} q_{s} N_{\nu}^{s}
   \,,
\end{align}
\end{subequations}
where $\Gamma^{\alpha}_{\mu \nu}$ are the Christoffel symbols, $\mathcal{C}_s$ is the collision operator and $N^{\mu}_{s}$ is the number density current
\begin{align}
   N^{\mu}_{s}
   \equiv
   \left<k^{\mu}\right>_{s}
   =
   \int d^4 k \sqrt{-g}  f_{s} k^{\mu}
   =
   \int \frac{d^3k}{\left|k_0\right|} \sqrt{-g} f_{s} k^{\mu}
   \,.
\end{align}

The first and second moments of the Vlasov equations yield the conservation laws for the number density and stress-tensor of species $s$
\begin{subequations}
\begin{align}
    \label{eq:particle-number-conservation-for-particle-s}
    &\nabla_{\alpha} \left( N_{s}^{\alpha} \right) = 0 \,, \\
    \label{eq:stress-energy-conservation-for-particle-s}
    &\nabla_{\mu} T^{\mu \nu}_{s} = q_{s} F^{\nu}{}_{\mu} N_{s}^{\mu} 
    \,,
\end{align}
\end{subequations}
where $T^{\mu \nu}_{s} \equiv m_{s} \left< k^{\mu} k^{\nu} \right>_{s}$. Given a timelike vector $U^{\mu}$, we decompose $N^{\mu}_{s}$ and $T^{\mu\nu}_{s}$ as
\begin{subequations}\label{eq:number-density-and-stress-tensor-s-def}
\begin{align}
    &N_{s}^{\mu} = n_{s} U^{\mu} + V^{\mu}_{s} \,,\\
    &T^{\mu \nu}_{s} = \mathcal{E}_{s} U^{\mu} U^{\nu} + \mathcal{P}_{s} \Delta^{\mu \nu} + 2 Q^{(\mu}_{s} U^{\nu)} + \pi^{\mu\nu}_{s} \,.
\end{align}
\end{subequations}
In terms of the moments of the distribution function, the decomposition is given by
\begin{subequations}\label{eq:moments-for-common-variables-species-s}
\begin{align}
    &n_{s} \equiv -N^{\mu}_{s} U_{\mu} = -\left< k_{\mu} U^{\mu}\right>_{s} \,,\\
    &V^{\mu}_{s} \equiv \Delta^{\mu}_{\nu} N^{\nu}_{s} = 
    \Delta^{\mu \nu}\left< k_{\nu} \right>_{s} 
    \,,\\
    &\mathcal{E}_{s} \equiv U_{\mu} U_{\nu} T^{\mu \nu}_{s} = m_s \left< (k_{\mu} U^{\mu})^2 \right>_{s}  \,,\\
    &\mathcal{P}_{s} \equiv \frac{1}{3} \Delta_{\mu \nu}  T^{\mu \nu}_{s} = \frac{1}{3} m_{s} \left< \Delta_{\mu \nu} k^{\mu} k^{\nu}\right>_{s}\,,\\
    &Q^{\mu}_{s} \equiv - \Delta^{\mu\beta} U^{\alpha} T_{\alpha \beta}^{s} = - m_{s} \left< \Delta^{\mu\beta} k_{\beta} U_{\alpha} k^{\alpha} \right>_{s} \,,\\
    & \pi^{\mu\nu}_{s} \equiv m_{s} \Delta^{\mu\nu}_{\alpha\beta}\left< k^{\mu} k^{\nu}\right>_{s}
    \,,
\end{align}
where $n_{s}$ is the number density, $V_{s}^{\mu}$ is the particle diffusion current, $\mathcal{E}_{s}$ is the energy density, $\mathcal{P}_{s}$ is the pressure, $\mathcal{Q}^{\mu}_{s}$ is the heat vector and $\pi_{s}^{\mu \nu}$ is the shear tensor.
The projection tensors $\Delta_{\mu\nu}$ and $\Delta^{\alpha \beta}_{\mu \nu}$ are defined as
\begin{align}
    &\Delta_{\mu\nu} \equiv g_{\mu \nu} + U_{\mu} U_{\nu} \,,\\
    &\Delta^{\alpha \beta}_{\mu \nu}
    \equiv
    \Delta^{(\alpha}_{\mu}\Delta^{\beta)}_{\nu} - \frac{1}{3}\Delta^{ \alpha \beta} \Delta_{\mu\nu}
    \,.
\end{align}
\end{subequations}

The fluid velocity vector $U^{\mu}$, which was used in Eq.~\eqref{eq:moments-for-common-variables-species-s}, has been arbitrary so far.
In this paper, we adopt the Eckhart frame to define $U^{\mu}$
\begin{align}\label{eq:Eckhart-frame-choice}
    \sum_{s} m_{s} V^{\mu}_{s} = 0
    \,.
\end{align}
For a review of different hydrodynamic frame choices, please see~\citet{Cercignani2002,Bemfica2022,Kovtun2019}.

With the choice of the fluid frame specified, we can define the mass-averaged current and the stress-energy tensor of the system
\begin{subequations}
\begin{align}
    \label{eq:number-density-all-plasma-def}
    &m N^{\mu} = \sum_{s} m_s N_{s}^{\mu} = U^{\mu} \sum_{s} m_s n_s
    \equiv
    m n U^{\mu} \,,\\
    &T^{\mu \nu}
    =
    \mathcal{E} U^{\mu} U^{\nu} + \mathcal{P} \Delta^{\mu \nu} + 2 Q^{(\mu} U^{\nu)} + \pi^{\mu\nu} \,,\\
    &\mathcal{E} = \sum_{s} \mathcal{E}_{s} = \sum_{s} m_s \left< (k_{\mu} U^{\mu})^2 \right>_{s}  \,,\\
    &\mathcal{P} = \sum_{s} \mathcal{P}_{s} = \frac{1}{3} \sum_{s} m_{s} \left< \Delta_{\mu \nu} k^{\mu} k^{\nu}\right>_{s}\,,\\
    &Q^{\mu} = \sum_{s} \mathcal{Q}^{\mu}_{s} = - \sum_{s} m_{s} \left< \Delta^{\mu\beta} k_{\beta} U_{\alpha} k^{\alpha} \right>_{s} \,,\\
    & \pi^{\mu\nu} = \sum_{s} \pi^{\mu\nu}_{s} = \sum_{s} m_{s} \Delta^{\mu\nu}_{\alpha\beta}\left< k^{\mu} k^{\nu}\right>_{s}
    \,,
\end{align}
\end{subequations}
where $m \equiv \sum_{s} m_{s}$ is a reference mass scale, equal to the sum of the particle masses of the species included in the plasma.
The evolution of the mean density current and stress energy tensor can be obtained by summing Eqs.~\eqref{eq:particle-number-conservation-for-particle-s} and \eqref{eq:stress-energy-conservation-for-particle-s} over all the particles
\begin{subequations}
\begin{align}
    \label{eq:conservation-number-density-all-species}
    &\nabla_{\alpha}\left(N^{\alpha} \right) = \nabla_{\alpha} \left( n U^{\alpha} \right) = 0\,,\\
    &\nabla_{\alpha} \left( T^{\mu \nu} + T^{\mu \nu}_{\mathrm{EM}} \right) = 0\,,
\end{align}
\end{subequations}
where the electromagnetic stress energy tensor is 
\begin{align}
    T^{\mu \nu}_{\mathrm{EM}} \equiv \frac{1}{4\pi}
    \bigg(F^{\mu \lambda} F^{\nu}{}_{\lambda} - \frac{1}{4} g^{\mu\nu} F^{\alpha \beta} F_{\alpha \beta} 
    \bigg) \,.
\end{align}
Observe that Eqs.~\eqref{eq:number-density-all-plasma-def} and \eqref{eq:conservation-number-density-all-species} are only valid in the Eckhart frame where particle diffusion over all the species in the plasma vanishes [Eq.~\eqref{eq:Eckhart-frame-choice}].
\subsection{Kinetic MHD}\label{sec:KMHD-section}
Let us consider the characteristic scaling of each of the terms in Eq.~\eqref{eq:Vlasov-Maxwell-Eqn} in a locally Minkowskian frame.
Let $L_{c}$ denote the characteristic scale of spatiotemporal variation at the macroscopic level, $\tau_{c}$ the collision time scale, $R_{c}$ the characteristic radius of particle motion, and $\omega_{c}$ the characteristic frequency.
With these definitions, we can obtain scaling of the different terms in Eq.~\eqref{eq:Vlasov-Maxwell-Eqn}
\begin{subequations}
\begin{align}
    &\left|k^{\mu} \frac{\partial f_{s}}{\partial x^{\mu}} \right|
    \sim \frac{|k| |f_s|}{L_{c}}
    \sim 
    \frac{R_c \omega_c |f_s|}{L_{c}}
    \,,\\
    &\left[\frac{q_{s}}{m_{s}} F^{\alpha}{}_{\mu} k^{\mu}\right]\frac{\partial f_{s}}{\partial k^{\alpha}}
    \sim
    \frac{\left|a_{\mathrm{particle}}\right|}{|k|} |f_s|
    \sim \frac{R_c \omega_c^2 |f_s| }{|k|}
    \sim 
    \omega_c |f_s| \,,\\
    &\mathcal{C}_s[f_s]
    \sim \tau_{c}^{-1} |f_s|
    \,.
\end{align}
\end{subequations}

In this paper, we are interested in capturing the dynamics of diffuse accretion flow around a SMBH.
Consider an accretion flow around a SMBH of mass $M_{\mathrm{BH}}$, with a characteristic magnetic field $B_c$. 
For simplicity, we assume the charged particles in the plasma are electrons and protons. In typical hot and diffusive black hole accretion flows, $L_c \sim M_{\mathrm{SMBH}}$; the characteristic radius for particle motion is the Larmor radius, the characteristic frequency is the cyclotron frequency, and the collisional time scale is set by Coulomb collisions. For an M87-like system~\citep{EHT-2021}, we obtain,
\begin{subequations}
\begin{align}
    &L_c \sim M_{\mathrm{SMBH}} \sim 1.5 \times 10^{11} \mathrm{m} \left(\frac{M}{10^8 M_{\odot}} \right) \,,\\
    &R_c \sim \frac{m_{p} v_{\mathrm{th}}}{e B_c}
    \sim 521 \mathrm{m} \,\left(\frac{v_{\mathrm{th}}}{c}\right) \left(\frac{30 \, \mathrm{G}}{B_c} \right)\,,\\
    &\omega_c \sim \frac{e B_c}{m_{p}} \sim 
    2.8 \times 10^5 \mathrm{Hz} \left(\frac{B_c}{30 \, \mathrm{G}}\right)
    \,,\\
    &\tau_c^{-1} \sim \frac{v_{\mathrm{th}}}{n_{\mathrm{e,i}} \lambda_{D}^4} \sim 
    1.32 \times 10^{-6} \mathrm{Hz}
    \,
    \left(\frac{v_{\mathrm{th}}}{c}\right) 
    \left(\frac{n_{e,i}}{10^{7} \mathrm{cm}^{-3}}\right)
    \left(\frac{T_{e,i}}{10^{10} \mathrm{K}}\right)^{-2}
    \,,
\end{align}
\end{subequations}
where $\lambda_D$ is the Debye length.
Hence, for diffusive accretion flows around a SMBH, we can safely assume that
\begin{align}\label{eq:ordering-check}
    \omega_c \gg \frac{R_c \omega_c}{L_c} > \tau_c^{-1} \,,
\end{align}
which implies that
\begin{align}\label{eq:collisionless-condition}
    \left|\left[\frac{q_{s}}{m_{s}} F^{\alpha}{}_{\mu} k^{\mu}\right]\frac{\partial f_{s}}{\partial k^{\alpha}}
    \right|
    \gg 
    \left|k^{\mu} \frac{\partial f_{s}}{\partial x^{\mu}} \right|
    >
    \mathcal{C}_s[f_s]
    \,.
\end{align}

We now implement the ordering obtained in Eq.~\eqref{eq:collisionless-condition} formally.
Consider an expansion of the form~\citep{1983bpp..conf....1K}
\begin{align}\label{eq:formal-expansion-ansatz}
    f_{s} = \sum_{n=0}^{\infty} f_{n,s} \epsilon^{-n} \,,
\end{align}
where $\epsilon \sim \mathcal{O}\left(\left|L_c/R_c\right|\right)$.
At leading order, the Vlasov equation simplifies to
\begin{align}\label{eq:vlasov-zeroth-order-GR}
    F^{\alpha}{}_{\mu} k^{\mu} \frac{\partial f_{s,0}}{\partial k^{\alpha}} = 0\,.
\end{align}
We show below that this condition implies the distribution function is gyrotropic, as in the non-relativistic limit~\citep{1983bpp..conf....1K}.
Next, we analyze the stress-energy conservation equation [Eq.~\eqref{eq:stress-energy-conservation-for-particle-s}] for particle $s$.
To leading order, the Lorentz-force term on the right-hand side of Eq.~\eqref{eq:stress-energy-conservation-for-particle-s} dominates, so that each species satisfies
\begin{align}\label{eq:degeneracy-condition}
    F^{\nu}{}_{\mu}N^{\mu}_{s}
    =
    0
    \,.
\end{align}
Multiplying this equation by $m_s$, summing over species, and using Eq.~\eqref{eq:number-density-all-plasma-def}, we obtain
\begin{align}
    F^{\nu}{}_{\mu}
    \sum_s m_s N^{\mu}_{s}
    =
    mnF^{\nu}{}_{\mu}U^{\mu}
    =
    0
    \,.
\end{align}
Thus,
\begin{align}
    F^{\nu}{}_{\mu}U^{\mu}
    =
    0
\end{align}
is a leading-order degeneracy condition of the KMHD expansion. 
It does not imply that the comoving electric field vanishes at all orders. In particular, the subleading parallel electric field $E_{\parallel}$ enters the gyroaveraged drift-kinetic equation and the corresponding moment equations; see Eqs.~\eqref{eq:gyroaveraged-kinetic-equation} and \eqref{eq:fluid-eom-species-s} below.

Finally, the electromagnetic field equations [Eq.~\eqref{eq:EM-equations}] imply charge neutrality,
\begin{align}\label{eq:charge-neutrality}
    \sum_s q_s N^{\mu}_{s}
    =
    0
    \implies
    \sum_s q_s n_s
    =
    0
    \,,
    \qquad
    \sum_s q_s V^{\mu}_{s}
    =
    0
    \,.
\end{align}
To show that the plasma is gyrotropic [Eq.~\eqref{eq:gyrotropic-plasma-condition}], we need to perform a coordinate transformation for the velocity variables.
In the non-relativistic theory~\citep{1983bpp..conf....1K}, one decomposes the velocity $k^{\mu}$ parallel and perpendicular to the magnetic field.
In relativity, we use the fact that electrogmagnetic field tensor is magnetically dominated [Eq.~\eqref{eq:degeneracy-condition}] to pick an orthonormal basis $(U^{\mu}, \hat{X}^{\mu}, \hat{Y}^{\mu}, \hat{b}^{\mu})$ and decompose the velocity vector $k^{\mu}$ as (see, Appendix B.6~\cite{Chael2023})
\begin{align}\label{eq:kmu-decomposition}
    k^{\mu} = W_{U} U^{\mu} + v_{\perp} \cos(\vartheta) \hat{X}^{\mu} + v_{\perp} \sin(\vartheta) \hat{Y}^{\mu} + v_{\parallel} \hat{b}^{\mu}\,,
\end{align}
where $b^{\mu}$ is the magnetic field in the fluid rest frame, $\vartheta$ is the gyroangle, $W_U$ is the Lorentz factor, $v_{\parallel}$ is the component of the velocity parallel to the magnetic field and $v_{\perp}$ is the component of the velocity perpendicular to the magnetic field.
The covariant decomposition of these quantities is
\begin{subequations}
\begin{align}
    &b^{\mu} \equiv U_{\nu} \left(\star F\right)^{\nu \mu} = \frac{1}{2} U_{\delta} \epsilon^{\delta \mu \alpha \beta } F_{\alpha \beta} \,, \,
    \hat{b}^{\mu} \equiv \frac{b^{\mu}}{|b|}
    \,,\, 
    W_{U} \equiv - k_{\mu} U^{\mu} = \sqrt{1 + v_{\perp}^2 + v_{\parallel}^2} \,,\\
    &v_{\parallel} \equiv k_{\mu} \hat{b}^{\mu} \,,
    v_{\perp} = \sqrt{\left(k_{\mu} \hat{Y}^{\mu}\right)^2 + \left(k_{\mu} \hat{X}^{\mu}\right)^2} \,,
    \tan \vartheta = \frac{\left(\hat{Y}^{\mu} k_{\mu} \right)}{\left(\hat{X}^{\mu} k_{\mu} \right)}\,.
\end{align}
\end{subequations}
As we show in Appendix~\ref{appendix:derivation-gyrokinetic}, one can change variables from $(x^{\mu},k) \to (x,v_{\parallel}, v_{\perp}, \vartheta)$ and show that Eq.~\eqref{eq:vlasov-zeroth-order-GR} reduces to
\begin{align}\label{eq:gyrotropic-plasma-condition}
    \frac{\partial f_{0,s}}{\partial \vartheta} = 0
    \,,
\end{align}
signifying that the plasma is gyrotropic\footnote{See~\citet{trent2023newcovariantformalismkinetic,2025arXiv250711616T,2025ApJ...987..101T} for a discussion on the drift-kinetic motion of test particles in curved spacetimes.}.

To understand the evolution of the distribution function in $(x,v_{\parallel}, v_{\perp})$ space, we continue the formal expansion of the distribution function[Eq.~\eqref{eq:formal-expansion-ansatz}] to the next order and average the Vlasov equation over $\vartheta$ (see Appendix~\ref{appendix:derivation-gyrokinetic} for details). The result is the relativistic generalization of the gyroaveraged kinetic equation
\begin{align}\label{eq:gyroaveraged-kinetic-equation}
    &W_{U} D_{\parallel} f_{0,s}
    +
    \frac{W_{U} v_{\perp}D_{\parallel} b }{2 b} 
    \frac{\partial f_{0,s}}{\partial v_{\perp}}
    +
    \frac{\partial f_{0,s}}{\partial v_{\parallel}}
    \bigg[
    \frac{q_s}{m_s} E_{\parallel} W_{U} 
    - W_{U}^2 \hat{b}^{\beta} D_{\parallel} U_{\beta}
    -  \frac{\hat{b}^{\alpha } v_{\perp}^2 \nabla_{\alpha }b}{2 b}
    \bigg]
    \nonumber\\
    &=
    \mathcal{C}_{s}\left[f_{s}\right]
    ,
\end{align}
where $E_{\parallel} \equiv F^{\mu \nu} U_{\nu} \hat{b}_{\mu}$ is the parallel electric field and $D_{\parallel}$ is the derivative along the gyroaveraged particle trajectory
\begin{align}
    D_{\parallel}
    \equiv
    D
    +
    \frac{v_{\parallel} \hat{b}^{\alpha}}{W_U}\nabla_{\alpha}
    =
    U^{\alpha}\nabla_{\alpha} + \frac{v_{\parallel} \hat{b}^{\alpha}}{W_U}\nabla_{\alpha}
    \,.
\end{align}
To obtain the non-relativistic limit, we can set $W_{U} = 1$ and see that Eq.~\eqref{eq:gyroaveraged-kinetic-equation} reduces to Eq. (20.90) of~\citet{notes-Alex}. 

The gyrotropic nature of the distribution function simplifies the number density current and the stress-energy tensor. 
Using Eqs.~\eqref{eq:gyrotropic-plasma-condition} in~\eqref{eq:number-density-and-stress-tensor-s-def}, one can show that for a gyrotropic plasma
\begin{subequations}\label{eq:gyrotropic-density-stress-components}
\begin{align}
    &n_{s} =  \left< W_{U} \right>_{s} \,, \\
    &V^{\mu}_{s} = \left< v_{\parallel} \right>_{s} \hat{b}^{\mu}
    \equiv V_{s} \hat{b}^{\mu}
    \,,\\
    &\mathcal{E}_{s} = m_{s} \left< W^2_{U} \right>_{s} \,, \\
    &\mathcal{P}_{s} = \frac{m_{s}}{3} \left<v_{\perp}^2 + v_{\parallel}^2\right>_{s}
    =
    \frac{2 p_{\perp,s} + p_{\parallel,s}}{3}
    \,,\\
    &\mathcal{Q}^{\mu}_{s} = m_{s} \left< W_{U} v_{\parallel} \right>_{s} \hat{b}^{\mu}
    \equiv \mathcal{Q}_{s} \hat{b}^{\mu}
    \,,\\
    \label{eq:pimunu-gyrotropic}
    &\pi^{\mu \nu}_{s}
    =
    m_{s} \left<v_{\parallel}^2 - \frac{v_{\perp}^2}{2} \right>_{s} \hat{b}^{<\mu} \hat{b}^{\nu>}
    =
    \left(p_{\parallel,s} - p_{\perp,s}\right) \hat{b}^{<\mu} \hat{b}^{\nu>}
    \,,
\end{align}
\end{subequations}
where the pressures parallel and perpendicular to the magnetic field are
\begin{align}
    p_{\parallel,s} \equiv m_s \left<v_{\parallel}^2\right>_s \,,
    \quad
    p_{\perp,s} \equiv \frac{m_s \left<v_{\perp}^2\right>_s}{2}\,.
\end{align}
Physically, Eq.~\eqref{eq:gyrotropic-density-stress-components} shows that particle and heat transport in the plasma ($V^{\mu}_{s}$ and $\mathcal{Q}^{\mu}_{s}$) are always along the magnetic field in the fluid rest frame. The difference between $p_{\parallel,s}$ and $p_{\perp,s}$ drives a shear stress in the plasma through $\pi^{\mu \nu}_{s}$. In the ideal MHD limit, $p_{\parallel,s} = p_{\perp,s}$ and there is no anisotropic stress in the plasma.
\subsection{Moments of the gyroaveraged equations}\label{sec:moment-equations}
Before giving the evolution equations for the moments of the gyrotropic distribution function, we introduce some notation.
We denote the most general moment of the gyrotropic distribution function as
\begin{align}\label{eq:IPQ-def}
    &\mathcal{I}^{(P,Q,R)}_{s} \equiv \left<  v_{\perp}^{P} v_{\parallel}^{Q} W_{U}^{R} \right>_{s} \nonumber\\
    &= 
    \int_{v_{\perp}=0}^{\infty}\int_{v_{\parallel}=-\infty}^{\infty}\int_{\vartheta=0}^{2\pi} \frac{v_{\perp} dv_{\perp} dv_{\parallel} d\vartheta}{\sqrt{1+v_{\perp}^2+v_{\parallel}^2}} f_{0,s} v_{\perp}^{P} v_{\parallel}^{Q} \left( 1+v_{\perp}^2+v_{\parallel}^2\right)^{R/2}
    \,.
\end{align}
For example, using the notation above, the number density, energy density, and pressure [Eq.~\eqref{eq:moments-for-common-variables-species-s}] are
\begin{align}
    n_{s} = - \mathcal{I}^{(0,0,1)}_{s} \,,
    \mathcal{E}_{s} = m_s \mathcal{I}^{(0,0,2)}_{s} \,,
    \mathcal{P}_{s} = \frac{m_s}{3} \left(\mathcal{I}^{(2,0,0)}_{s} + \mathcal{I}^{(0,2,0)}_{s}\right)
    \,.
\end{align}
We also introduce short hands for certain common moments of the distribution function that appear in our equations (for definitions in the non-relativistic limit see below Eq. (13) in~\citet{1997PhPl....4.3974S})
\begin{subequations}
\begin{align}
    &\mathcal{Q}_{\parallel,s}^{(-k)} \equiv m_s \mathcal{I}_{s}^{(0,3,-1-k)} \,, \quad
    \mathcal{Q}_{\perp,s}^{(-k)} \equiv \frac{1}{2} m_s \mathcal{I}_{s}^{(2,1,-1-k)}\,, \\
    &p_{\perp,s}^{(-k)} \equiv \frac{m_s \mathcal{I}_{s}^{(2,0,-k)}}{2} \,, \quad 
    p_{\parallel,s}^{(-k)} \equiv m_s \mathcal{I}_{s}^{(0,2,-k)} \,,\\
    &r_{\parallel,\perp,s}^{(-k)} \equiv \frac{1}{2}m_s \mathcal{I}_{s}^{(2,2,-k)} \,, \quad
    r_{\parallel,\parallel,s}^{(-k)} \equiv m_s \mathcal{I}_{s}^{(0,4,-k)} \,, \quad
    r_{\perp,\perp,s}^{(-k)} \equiv \frac{1}{4} m_s \mathcal{I}_{s}^{(4,0,-k)}\,.
\end{align}
\end{subequations}
In the above equations, when $k=0$, we drop the superscript to simplify notation. For example, we denote $\mathcal{Q}_{\parallel,s}^{(-0)}$ as $\mathcal{Q}_{\parallel,s}$.
Physically, $\mathcal{Q}_{\parallel,s}^{(-k)}$ is the (Lorentz factor weighted) heat flux parallel to the magnetic field, $\mathcal{Q}_{\perp,s}^{(-k)}$ is the (Lorentz factor weighted) heat flux perpendicular to the magnetic field, and $(r_{\parallel,\perp,s}^{(-k)}, r_{\parallel,\parallel,s}^{(-k)},  r_{\perp,\perp,s}^{(-k)})$ are Lorentz factor weighted fourth-order moments of the distribution function.

Let us now introduce operators that will be used to write down the evolution equations.
Define
\begin{subequations}
\begin{align}
\label{eq:Oparallel-def}
    &O_{\parallel} \left[\mathcal{I}^{(P,Q,R)}_s\right] \equiv \left<  v_{\perp}^{P} v_{\parallel}^{Q} W_{U}^{R} \frac{\partial \log f_{0,s} }{\partial {v_{\parallel}}}  \right>_{s}
    =
    - Q \mathcal{I}^{(P,Q-1,R)}_s - (R-1) \mathcal{I}^{(P,Q+1,R-2)}_s
    \,,\\
\label{eq:Operp-def}
    &O_{\perp} \left[\mathcal{I}^{(P,Q,R)}_s\right] \equiv \left<  v_{\perp}^{P} v_{\parallel}^{Q} W_{U}^{R} \frac{\partial \log f_{0,s} }{\partial {v_{\perp}}}  \right>_{s}
    =
    - (P+1) \mathcal{I}^{(P-1,Q,R)}_s - (R-1) \mathcal{I}^{(P+1,Q,R-2)}_s
    \,,\\
\label{eq:Oc-def}
    &O_{c}\left[\mathcal{I}_s^{(P,Q,R)}\right]
    \equiv
    \int
    \frac{v_{\perp} dv_{\perp} dv_{\parallel} d\vartheta}{\sqrt{1+v_{\perp}^2+v_{\parallel}^2}} \mathcal{C}_{s}\left[f_{0,s}\right]
    v_{\perp}^{P} v_{\parallel}^{Q} \left( 1+v_{\perp}^2+v_{\parallel}^2\right)^{R/2}
    \,.
\end{align}
\end{subequations}
The second equality in Eqs.~\eqref{eq:Oparallel-def} and \eqref{eq:Operp-def} is obtained by integrating by parts in $v_{\parallel}$ and $v_{\perp}$, respectively.
To obtain the evolution equation for the moment $\mathcal{I}_{s}^{(P,Q,R)}$, we multiply the drift-kinetic equation [Eq.~\eqref{eq:gyroaveraged-kinetic-equation}] by $ v_{\perp}^{P} v_{\parallel}^{Q} W_{U}^{R-1}$ and integrate over velocity space.
Before we present the results of this calculation, let us define
\begin{subequations}
\begin{align}
    &I_0 \equiv
    \left<D_{\parallel} f_{0,s} v_{\perp}^{P} v_{\parallel}^{Q} W_{U}^{R} \right>
    =
    D \mathcal{I}^{(P,Q,R)}_{s} + \hat{b}^{\alpha} \nabla_{\alpha} \mathcal{I}_{s}^{(P,Q+1,R-1)}
    \,,\\
    &I_1 \equiv
    \left<
    \frac{v_{\perp}^{P +1} v_{\parallel}^{Q} W_{U}^{R}D_{\parallel} b }{2 b} 
    \frac{\partial f_{0,s}}{\partial v_{\perp}} \right> 
    =
    \frac{D b}{2 b} O_{\perp} \left[\mathcal{I}^{(P+1,Q,R)}_{s}\right]
    +
    \frac{\hat{b}^{\alpha} \nabla_{\alpha} b}{2 b} O_{\perp} \left[\mathcal{I}^{(P+1,Q+1,R-1)}_{s}\right]
    \,,\\
    &I_2 \equiv
    \left<
    v_{\perp}^{P} v_{\parallel}^{Q} W_{U}^{R}
    \frac{\partial f_{0,s}}{\partial v_{\parallel}}
    \bigg[
    \frac{q_s}{m_s} E_{\parallel} 
    - W_{U} \hat{b}^{\beta} D_{\parallel} U_{\beta}
    -  \frac{\hat{b}^{\alpha } v_{\perp}{}^2 \nabla_{\alpha }b}{2 b W_{U}}
    \bigg]
    \right>
    \nonumber\\
    &=
    \frac{q_s}{m_s}
    E_{\parallel} O_{\parallel} \left[\mathcal{I}^{(P,Q,R)}_s\right]
    -
    \hat{b}^{\beta} D U_{\beta}
    O_{\parallel}  \left[\mathcal{I}^{(P,Q,R+1)}_s\right]
    -
    \hat{b}^{\alpha} \hat{b}^{\beta} \nabla_{(\alpha} U_{\beta)} O_{\parallel}  \left[\mathcal{I}^{(P,Q+1,R)}_s\right]
    \nonumber\\
    &-
    \frac{\hat{b}^{\alpha} \nabla_{\alpha} b}{2 b} 
    O_{\parallel}  \left[\mathcal{I}^{(P+2,Q,R-1)}_s\right]
    \,,
\end{align}
\end{subequations}
where $I_0$ represents the advection of the moments parallel to the velocity and magnetic fields and $I_1$, $I_2$ are the contributions from the velocity derivative of the distribution function perpendicular and parallel to the magnetic field.  
The moment equations are the combined sum of $I_0$, $I_1$, $I_s$ and the contributions obtained from collisions 
\begin{align}\label{eq:moment-equations-general}
    &I_0 + I_1 + I_2 = O_{c}\left[\mathcal{I}^{(P,Q,R-1)}\right]\,.
\end{align}
The equations above describe the evolution equation for the most general moment of the gyrotropic distribution function.
In this paper, we focus only on the evolution equation for the first few moments: the number density, the stress-energy tensor, the parallel pressure, and the perpendicular pressure.
These equations are
\begin{subequations}\label{eq:fluid-eom-species-s}
\begin{align}
    \label{eq:number-density-conservation-eqn-species-s}
    &\nabla_{\alpha} \left(n_{s} U^{\alpha} + V^{\alpha}_{s} \right) = 0\,,\\
    \label{eq:stress-energy-conservation-species-s}
    &\nabla_{\mu} T^{\mu \nu}_{s} = q_s F^{\nu}{}_{\mu} N^{\mu}_{s} \,,\\
    \label{eq:parallel-pressure-evolution-species-s}
    &D p_{\parallel,s}
    +
    \hat{b}^{\alpha} \nabla_{\alpha} \mathcal{Q}_{\parallel,s}
    -\frac{\hat{b}^{\alpha }\nabla _{\alpha }b \left(\mathcal{Q}_{\|, s}-2
   \mathcal{Q}_{\perp,s}\right)}{b}
   +
   \frac{Db
   r_{\|,\perp,s}^{(-2)}}{b}
   -\hat{b}^{\alpha }\hat{b}^{\beta }\nabla _{\alpha }U_{\beta
   } r_{\|,\|,s}^{(-2)}
   \nonumber\\
   &+E_{\|} q_s \left(\frac{\mathcal{Q}_{\|,s}^{(-1)}}{m_s}-2 V_s\right)
   +p_{\|,s}
   \left(3 \hat{b}^{\alpha }\hat{b}^{\beta }\nabla _{\alpha }U_{\beta
   }-\frac{Db}{b}\right)
   +
   2 Q_s \hat{b}^{\alpha }DU_{\alpha }
    =
    m_s O_{c}\left[ \mathcal{I}^{(0,2,-1)}_s \right]
    \,, \\
    \label{eq:perpendicular-pressure-evolution-species-s}
    &D p_{\perp,s}
    +
    \hat{b}^{\alpha} \nabla_{\alpha}\mathcal{Q}_{\perp,s}
    -\hat{b}^{\alpha }\hat{b}^{\beta }\nabla _{\alpha }U_{\beta }
    r_{\|,\perp,s}^{(-2)}
    +
    \frac{Db
    r_{\perp,\perp,s}^{(-2)}}{b}-
    \frac{2 \hat{b}^{\alpha }\nabla _{\alpha }b
    \mathcal{Q}_{\perp,s}}{b}
    +\frac{E_{\|} q_s
    \mathcal{Q}_{\perp,s}^{(-1)}}{m_s}
    \nonumber\\
    &+p_{\perp,s} \left(\hat{b}^{\alpha
    }\hat{b}^{\beta }\nabla _{\alpha }U_{\beta }-\frac{2 Db}{b}\right)
    =
    \frac{m_s}{2}O_{c}\left[ \mathcal{I}^{(2,0,-1)}_s \right]
    \,.
\end{align}
\end{subequations}
The moments required for closure of the above equations are listed in Table~\ref{tab:moments-for-closure}.
We emphasize that Eq.~\eqref{eq:stress-energy-conservation-species-s} retains the Lorentz force on each species. 
Although the leading-order KMHD ordering implies $F^{\nu}{}_{\mu}U^{\mu}=0$ and a vanishing leading-order perpendicular current, it does not imply that the full current vanishes. 
Indeed, summing Eq.~\eqref{eq:stress-energy-conservation-species-s} over species gives
\begin{align}
    \nabla_{\mu}T^{\mu\nu}
    =
    F^{\nu}{}_{\mu}J^{\mu}
    \,,
    \qquad
    J^{\mu}
    =
    \sum_s q_s N_s^{\mu}
    \,.
\end{align}
The first non-vanishing perpendicular current arises at subleading order in the KMHD expansion and provides the usual MHD Lorentz force, see Eq.~\eqref{eq:all-species-equations-3+1} below.

In the non-relativistic limit, the characteristic particle velocity $v_{\rm ch}$ is much less than the speed of light $c$.
The higher-order moments, such as $r^{(-2)}_{\parallel,\parallel,s}$, that appear in Eqs.~\eqref{eq:parallel-pressure-evolution-species-s} and \eqref{eq:perpendicular-pressure-evolution-species-s} are suppressed relative to the pressure moments by additional powers of $v_{\rm ch}/c$, where $v_{\rm ch}$ is the characteristic flow velocity.
For example, restoring factors of $c$,
\begin{align}
    r^{(-2)}_{\parallel,\parallel,s}
    =
    m_s
    \left\langle
    \frac{v_{\parallel}^4}{c^2}
    W_U^{-2}
    \right\rangle_s
    \sim
    m_s
    \left\langle
    \frac{v_{\parallel}^4}{c^2}
    \right\rangle_s
    \,,
\end{align}
where we have used $W_U^{-2}=1+\mathcal{O}(v_{\rm ch}^2/c^2)$.
Since
\begin{align}
    p_{\parallel,s}
    \sim
    m_s
    \left\langle
    v_{\parallel}^2
    \right\rangle_s
    \,,
\end{align}
we have
\begin{align}
    \frac{r^{(-2)}_{\parallel,\parallel,s}}{p_{\parallel,s}}
    \sim
    \mathcal{O}
    \left(
    \frac{v_{\rm ch}^2}{c^2}
    \right)
    \,.
\end{align}
Thus, the appearance of higher powers of the particle velocity in the negative-Lorentz-factor-weighted moments makes these moments higher order in the non-relativistic expansion.
Neglecting such higher-order moments and setting $\mathcal{Q}=V=0$ in Eq.~\eqref{eq:parallel-pressure-evolution-species-s} results in equations identical to Eqs.~(16) and (17) of~\citet{1997PhPl....4.3974S}.
\begin{table*}
    \centering
\begin{tabular}{c c }
    \hline
        \textbf{Description} & \textbf{Moment } \\
    \hline
     Heat flux and particle diffusion & $\mathcal{Q}_s$ and $V_s$ \\
     \hline
     Parallel heat flux & $\left\{
        \mathcal{Q}_{\parallel,s},
        \mathcal{Q}_{\parallel,s}^{(-1)}
     \right\}$  \\
     \hline
    Perpendicular heat flux & $\left\{
        \mathcal{Q}_{\perp,s},
        \mathcal{Q}_{\perp,s}^{(-1)}
     \right\}$  \\
     \hline
     Lorentz factor weighted pressure moments & 
     $\left\{
     r_{\perp,\perp,s}^{(-2)}, r_{\parallel,\parallel,s}^{(-2)}, r_{\parallel,\perp,s}^{(-2)}
     \right\}$  \\
     \hline
\end{tabular}
\caption{List of the moments needed to close Eq.~\eqref{eq:fluid-eom-species-s}.}
    \label{tab:moments-for-closure}
\end{table*}

We obtain the evolution equation for the species-averaged flow from Eq.~\eqref{eq:fluid-eom-species-s} by summing over $s$
\begin{subequations}\label{eq:all-species-equations-3+1}
\begin{align}
    \label{eq:number-density-conservation-all-species}
    &\nabla_{\alpha} \left(n U^{\alpha} \right) = 0\,,\\
    \label{eq:stress-energy-conservation-all-species}
    &\nabla_{\mu} \left(T^{\mu \nu} + T^{\mu \nu}_{\mathrm{EM}}\right) = 0 \,,\\
    \label{eq:pparallel-Eqn-all-species}
    &D p_{\parallel}
    +
    \hat{b}^{\alpha} \nabla_{\alpha} \mathcal{Q}_{\parallel}
    -\frac{\hat{b}^{\alpha }\nabla _{\alpha }b \left(\mathcal{Q}_{\|}-2
   \mathcal{Q}_{\perp}\right)}{b}
   +
   \frac{Db
   r_{\|,\perp}^{(-2)}}{b}
   -\hat{b}^{\alpha }\hat{b}^{\beta }\nabla _{\alpha }U_{\beta
   } r_{\|,\|}^{(-2)}
   \nonumber\\
   &+E_{\|} \sum_{s} q_s \left(\frac{\mathcal{Q}_{\|,s}^{(-1)}}{m_s}\right)
   +p_{\|}
   \left(3 \hat{b}^{\alpha }\hat{b}^{\beta }\nabla _{\alpha }U_{\beta
   }-\frac{Db}{b}\right)
   +
   2 \mathcal{Q} \hat{b}^{\alpha }DU_{\alpha }
    =
    \sum_{s} m_s O_{c}\left[ \mathcal{I}^{(0,2,-1)}_s \right]
    \,, \\
    \label{eq:pperp-Eqn-all-species}
    &D p_{\perp}
    +
    \hat{b}^{\alpha} \nabla_{\alpha}\mathcal{Q}_{\perp}
    -\hat{b}^{\alpha }\hat{b}^{\beta }\nabla _{\alpha }U_{\beta }
   r_{\|,\perp}^{(-2)}
   +
   \frac{Db
   r_{\perp,\perp}^{(-2)}}{b}-
   \frac{2 \hat{b}^{\alpha }\nabla _{\alpha }b
   \mathcal{Q}_{\perp}}{b}
   +E_{\|}
   \sum_{s}\frac{ q_s
   \mathcal{Q}_{\perp,s}^{(-1)}}{m_s}
   \nonumber\\
   &+p_{\perp} \left(\hat{b}^{\alpha
   }\hat{b}^{\beta }\nabla _{\alpha }U_{\beta }-\frac{2 Db}{b}\right)
    =
    \sum_{s}\frac{m_s}{2}O_{c}\left[ \mathcal{I}^{(2,0,-1)}_s \right]
    \,.
\end{align}
\end{subequations}

We now analyze the above equations to understand their implications.
Equations~\eqref{eq:conservation-number-density-all-species} and \eqref{eq:stress-energy-conservation-all-species} provide the conservation laws for the number density and the stress-energy tensor averaged over all the species in the plasma.
The stress-energy tensor [Eq.~\eqref{eq:gyrotropic-density-stress-components}] depends on the parallel pressure, perpendicular pressure, and heat flux. The evolution of the parallel and perpendicular pressures is given by Eqs.~\eqref{eq:pparallel-Eqn-all-species} and \eqref{eq:pperp-Eqn-all-species}, which contain unknown moments listed in Table~\ref{tab:moments-for-closure}.
In Sec.~\ref{sec:Landau-Closure}, we analyze possible closures to consistently evolve the equations above in the presence of heat fluxes. For now, we ignore the heat fluxes and show that our equations reduce to the familiar CGL equations.
\subsection{Double adiabatic limit}\label{sec:double-adiabatic-limit}
Consider a simple closure for Eq.~\eqref{eq:all-species-equations-3+1}~\citep{CGL-Original} and ignore the contributions of collisions, heat fluxes, and Lorentz factor weighted pressure moments in Table~\ref{tab:moments-for-closure}.
With these assumptions, the stress-energy tensor does not contain contributions from the heat flux, and Eqs.~\eqref{eq:pparallel-Eqn-all-species} and \eqref{eq:pperp-Eqn-all-species} reduce to
\begin{subequations}\label{eq:CGL-1}
\begin{align}
    &D p_{\parallel}
    +p_{\|}
    \left(3 \hat{b}^{\alpha }\hat{b}^{\beta }\nabla _{\alpha }U_{\beta
    }-\frac{Db}{b}\right)
    =
    0
    \,, \\
    &D p_{\perp}
   +p_{\perp} \left(\hat{b}^{\alpha
   }\hat{b}^{\beta }\nabla _{\alpha }U_{\beta }-\frac{2 Db}{b}\right)
    =
    0\,.
\end{align}
\end{subequations}
Using $b_{\beta} \nabla_{\alpha} \left(\star F^{\alpha \beta} \right) = 0$ and $\nabla_{\mu}\left(n U^{\mu}\right) = 0$, one can show that
\begin{align}
    \hat{b}^{\alpha }\hat{b}^{\beta }\nabla _{\alpha }U_{\beta
    } = \frac{Db}{b} - \frac{Dn}{n}
    \,.
\end{align}
Hence, Eq.~\eqref{eq:CGL-1} reduces to
\begin{subequations}\label{eq:CGL-2}
\begin{align}
    &\frac{n^3}{b^2}
    D \left(\frac{p_{\parallel} b^2}{n^3}\right)
    =
    0
    \,, \\
    &n b D \left(\frac{p_{\perp}}{n b} \right)
    =
    0\,.
\end{align}
\end{subequations}
The above equations generalize the CGL equations of non-relativistic MHD to relativity~\citep{Gedalin-1995}. However, there is an important assumption underlying their derivation.
While one can neglect the heat fluxes in Table~\ref{tab:moments-for-closure}, one cannot always ignore the Lorentz factor weighted pressure moments for all equations of state. For non-relativistic flows and equations of state, these contributions can be safely ignored. For ultra-relativistic flows, however, these moments are of the same order as the parallel pressure moments and cannot be ignored (see Sec.~\ref{sec:Landau-Closure}).
\subsection{Summary of the important equations}\label{sec:KMHD-summary}
The evolution of gyrotropic plasma in general relativity is described by the following equations:
\begin{itemize}
    \item Eckhart frame choice [Eq.~\eqref{eq:Eckhart-frame-choice}]
    \begin{align}\label{eq:Eckhart-frame-choice-summary}
        \sum_{s} m_{s} V_{s} = 0\,.
    \end{align}
    \item Conservation of laws for the averaged number density, stress-energy tensor across all species  [Eqs.~\eqref{eq:number-density-conservation-all-species} and \eqref{eq:stress-energy-conservation-all-species}] and Gauss-Farday law [Eq.~\eqref{eq:Gauss-Farday-laws}]
        \begin{align}\label{eq:all-species-plus-maxwell-summary}
            \nabla_{\alpha} \left(n U^{\alpha} \right) = 0\,,\quad
            \nabla_{\mu} \left(T^{\mu \nu} + T^{\mu\nu}_{\mathrm{EM}}\right) = 0
            \,,
            \quad
            \nabla_{\mu} (\star F^{\mu \nu}) = 0 \,.
        \end{align}
    \item A kinetic equation describing the evolution of the gyrotropic distribution function [Eq.~\eqref{eq:gyroaveraged-kinetic-equation}]\footnote{Notice that we have dropped the subscript label $0$ from $f_{0,s}$ in Eq.~\eqref{eq:gyroaveraged-kinetic-equation-summary} to reduce clutter.}
        \begin{align}\label{eq:gyroaveraged-kinetic-equation-summary}
            &W_{U} D_{\parallel} f_{s}
            \!+\!
            \frac{W_{U} v_{\perp}D_{\parallel} b }{2 b} 
            \frac{\partial f_{s}}{\partial v_{\perp}}
            \!+\!
            \frac{\partial f_{s}}{\partial v_{\parallel}}
            \bigg[
            \frac{q_s}{m_s} E_{\parallel} W_{U} 
            - W_{U}^2 \hat{b}^{\beta} D_{\parallel} U_{\beta}
            -  \frac{\hat{b}^{\alpha } v_{\perp}^2 \nabla_{\alpha }b}{2 b}
            \bigg]
            \!
            =
            \!
            \mathcal{C}_{s}\left[f_{s}\right]\,.
        \end{align}
    \item Charge neutrality [Eq.~\eqref{eq:charge-neutrality}]
    \begin{align}\label{eq:charge-neutrality-summary}
        \sum_{s} q_{s} n_s = 0 \,, \quad \sum_{s} q_{s} V_{s} = 0\,.
    \end{align}
\end{itemize}
The kinetic equation derived directly couples to the macroscopic fluid equations.
The general evolution equations for the moments of the gyroaveraged distribution function are listed in Eq.~\eqref{eq:moment-equations-general}. Simplified evolution equations for the parallel and perpendicular pressures for species $s$ are given in Eqs.~\eqref{eq:parallel-pressure-evolution-species-s} and \eqref{eq:perpendicular-pressure-evolution-species-s}.
The evolution equations for the parallel and perpendicular moments, averaged across species and given in Eqs.~\eqref{eq:pparallel-Eqn-all-species} and \eqref{eq:pperp-Eqn-all-species}, provide a practical approach to evolving multi-species relativistic plasmas.
The higher-order moments needed to close these equations are provided in Table~\ref{tab:moments-for-closure}.
\section{Linearized KMHD equations in the absence of collisions}\label{sec:linear-response}
The discussion in the previous section was for a general plasma, with no assumptions about the number of species. Henceforth, we assume that the plasma consists of two equal and oppositely charged particles. The positively and negatively charged species will be called ions and electrons, respectively. In such a two-species plasma, the charge neutrality condition [Eq.~\eqref{eq:charge-neutrality}] implies that
\begin{align}
    n_e = n_i \,, \quad 
    V_{e} = V_{i} \,.
\end{align}
We can combine this with the Eckhart frame choice [Eq.~\eqref{eq:Eckhart-frame-choice}] to show that particle diffusion vanishes for the electron and ions as in the non-relativistic theory~\citep{1997PhPl....4.3974S}
\begin{align}\label{eq:V-zero-ion-electron}
    V_{e} = V_{i} = 0\,.
\end{align}

In this section, we analyze the linearized KMHD equations in Fourier space for an ion-electron plasma in the absence of collisions in Minkowski spacetime. We describe our choice for the equilibrium distribution function in Sec.~\ref{sec:anisotropic-equilibrium}.
We then focus on deriving and examining the linearized evolution equations for plasma moments in Sec.~\ref{sec:linear-response-KMHD}.
Finally, we list the analytical expressions for the moments obtained using KMHD in the high-temperature and low-frequency limit in Sec.~\ref{sec:KT-heat-fluxes-linearized}.
In the following subsections, we only present the most important equations and relegate technical calculations to Appendix~\ref{appendix:moments-of-distribution-function} and \ref{appendix:plasma-response-function}.
\subsection{Equilibrium distribution function}\label{sec:anisotropic-equilibrium}
Let us suppose that we are studying a gyrotropic plasma at rest, threaded by a constant magnetic field and a vanishing electric field in Minkowski space.
We set up our coordinate system so the fluid velocity aligns with the time direction and the magnetic field points along the $z$-direction
\begin{align}\label{eq:equilibrium-velocity-magnetic-field}
    U^{\mu} = (1,0,0,0)\,, \quad 
    \hat{b}^{\mu} = (0,0,0,1)\,.
\end{align}
A global equilibrium state of the KMHD equations [Eq.~\eqref{eq:gyroaveraged-kinetic-equation-summary}] without a parallel electric field is any function $f_{s}(v_{\parallel}, v_{\perp})$.
In non-relativistic theory, a common choice for the equilibrium distribution $f_{0,s}(v_{\parallel}, v_{\perp})$ is a bi-Maxwellian distribution function (see Eq. (23) of~\citet{1997PhPl....4.3974S})
\begin{align}\label{eq:fNewt-bi-Maxwellian}
    f_{s}^{\mathrm{non-rel}}
    \propto
    \exp\bigg[ 
    -\frac{m_s v_{\parallel}^2}{2 T_{\parallel,s}}
    -\frac{m_s v_{\perp}^2}{2 T_{\perp,s}}
    \bigg]\,,
\end{align}
where $T_{\parallel,s}$ and $T_{\perp,s}$ are the temperatures parallel and perpendicular to the magnetic field. A bi-Maxwellian distribution function captures pressure anisotropies, and one can analyze the linearized equations to obtain a closure for the heat fluxes~\citep{1997PhPl....4.3974S}.

However, it is important to note that a bi-Maxwellian distribution function in relativity does not reduce to the equilibrium Maxwell-Juttner distribution function in the absence of pressure anisotropies.
Therefore, we choose a distribution function inspired by anisotropic relativistic hydrodynamics~\citep{Alqahtani_2018}
\begin{align}\label{eq:anisotropic-distribution-function-relativity}
    f_{s}(v_{\parallel}, v_{\perp}) = \alpha_s \exp\left(-z_s \sqrt{1 + v_{\parallel}^2 \delta_{1,s} + v_{\perp}^2 \delta_{2,s}} \right)
    \,,
\end{align}
where $\alpha_{s}$ is a constant, $z_{s} \sim m_{s}/T_{s}$ quantifies the ratio of the rest mass energy to the thermal energy $(T_s)$ of particle, and $\delta_{1,s}\sim T_{s}/T_{\parallel,s}$ ($\delta_{2,s}\sim T_{s}/T_{\perp,s}$) quantifies the ratio of the total temperature to  the temperature parallel (perpendicular) to the magnetic field.
In the non-relativistic limit, Eq.~\eqref{eq:anisotropic-distribution-function-relativity} reduces to the bi-Maxwellian distribution function [Eq.~\eqref{eq:fNewt-bi-Maxwellian}]. 

Apart from the moments $\mathcal{I}_{s}^{(P,Q,R)}$ [Eq.~\eqref{eq:moment-equations-general}], we will also need to evaluate the following moments of $f_s$
\begin{align}\label{eq:Iminus-moment}
    \mathcal{I}^{(-1)}_{(P,Q,R),s} \equiv \left<\frac{1}{\tilde{W}} v_{\perp}^{P} v_{\parallel}^{Q} W^{R}  \right>_s
\end{align}
where $\tilde{W} \equiv \sqrt{1 + v_{\parallel}^2 \delta_{1,s} + v_{\perp}^2 \delta_{2,s}}$.
Strategies for numerically obtaining the moments $\mathcal{I}^{(P,Q,R)}_s$ and $\mathcal{I}^{(-1)}_{(P,Q,R),s}$ for general plasma temperature are in Appendix~\ref{appendix:moments-of-distribution-function}. In the high-temperature limit, the moments can be evaluated analytically; see Appendix~\ref{appendix:analytical-moments} for explicit expressions for the first few moments of the distribution function.
\subsection{Linearized equations}\label{sec:linear-response-KMHD}
The linearized gyroaveraged kinetic equation [Eq.~\eqref{eq:gyroaveraged-kinetic-equation-summary}] in the absence of collisions is
\begin{align}\label{eq:linearized-kinetic-equation}
    D_{\parallel} \delta f_s
    +
    \frac{v_{\perp} D_{\parallel} \delta b }{2 b} 
    \frac{\partial f_s}{\partial v_{\perp}}
    +
    \frac{\partial f_s}{\partial v_{\parallel}}
    \bigg[
    \frac{q_s}{m_s} \delta E_{\parallel} 
    - W_{U} \hat{b}^{\beta} D_{\parallel} \delta U_{\beta}
    -  \frac{v_{\perp}^2 \hat{b}^{\alpha} \nabla_{\alpha } \delta b}{2 b W_{U}}
    \bigg]
    = 0
    \,.
\end{align}
Here, $\delta U^{\mu}$ denotes the dynamical fluid four-velocity perturbation defined by our choice of Eckhart fluid frame. 
This arises due to the changes in the bulk-motion of the fluid 4-velocity due to inertia and magnetic pressure.
Its equilibrium value is provided in Eq.~\eqref{eq:equilibrium-velocity-magnetic-field}. To prevent clutter, we shall use $U^{\mu}$ to refer to its equilibrium value and $\delta U^{\mu}$ to denote the perturbation. Note that the velocity-space variables $v_{\parallel}$ and $v_{\perp}$ are defined with respect to the corresponding local fluid tetrad and remain well defined under this perturbation.

Now, we Fourier transform
\footnote{The formally correct way of defining this operation is to use a Laplace transform. This distinction is not relevant for the discussions in this article.}
in space and time, and replace 
\begin{align}
    \partial_t \to - i \omega \,,
    \partial_x \to i k_x \,,
    \partial_y \to i k_y \,,
    \partial_z \to i k_z \,.
\end{align}
In this section, $k_{x,y,z}$  represent the wave numbers for the perturbations and should not be confused with the particle $4$-velocity in the Vlasov-Maxwell equations.
In Fourier space, we simplify Eq.~\eqref{eq:linearized-kinetic-equation} as
\begin{align}\label{eq:linearized-kinetic-v1}
    \delta f_{s} = 
    - \frac{v_{\perp} \delta b}{2 b} \frac{\partial f_s}{\partial v_{\perp}} 
    +
    W_U \hat{b}^{\beta} \delta U_{\beta} 
    \frac{\partial f_s}{\partial v_{\parallel}} 
    -
    \frac{1}{D_{\parallel}}
    \frac{\partial f_s}{\partial v_{\parallel}}
    \bigg[
    \frac{q_s}{m_s} \delta E_{\parallel} 
    \bigg]
    +
    \frac{ v_{\perp}^2 }{ W_{U} D_{\parallel}}
    \frac{i k_z \delta b }{2 b}
    \frac{\partial f_s}{\partial v_{\parallel}}
    \,,
\end{align}
where in Fourier space
\begin{align}
    D_{\parallel} = i \left(- \omega + \frac{v_{\parallel}}{W_U} k_z\right) \,.
\end{align}
Integrating Eq.~\eqref{eq:linearized-kinetic-v1} over velocity space gives the linearized moments of the system
\begin{align}\label{eq:moments-from-kinetic-theory}
    \delta \mathcal{I}^{(M,N,Q)}_s
    &=
    O_{\parallel}\left[ \mathcal{I}^{(M,N,Q+1)}_s \right]
    \hat{b}^{\beta} \delta U_{\beta}
    -\frac{\delta b}{2 b}
    O_{\perp}\left[ \mathcal{I}^{(M+1,N,Q)}_s \right]
    -
    \frac{q_s \mathrm{sgn}^{N}(k_z)}{m_s (i k_z)} \delta E_{\parallel} 
    \mathcal{H}^{(M,N,Q)}_{s}
    \nonumber\\
    &+
    \frac{\mathrm{sgn}^N(k_z) \delta b}{2 b}
    \mathcal{H}^{(M+2,N,Q-1)}_{s}
    \,,
\end{align}
where $\mathrm{sgn}(\cdot)$ is the sign function
\begin{align}
    \zeta \equiv \frac{\omega}{|k_z|} \,,
\end{align}
and
\begin{align}\label{eq:Hfunc-definition}
    \mathcal{H}^{(P,Q,R)}_{s}(\zeta)
    \equiv
    \left<\frac{v_{\perp}^{P} v_{\parallel}^{Q} W_U^{R} }{v_{\parallel} W_{U}^{-1} - \zeta}\frac{\partial \log(f_s)}{\partial v_{\parallel}} \right>_s
    =
    -z_s \delta_{1,s}
    \left<\frac{v_{\perp}^{P} v_{\parallel}^{Q+1} W_U^{R} }{ \left(v_{\parallel} W_{U}^{-1} - \zeta\right) \Tilde{W}} \right>_s
    \,.
\end{align}
We have used Eq.~\eqref{eq:anisotropic-distribution-function-relativity} to obtain the second equality. We note that the linearized plasma must have zero particle diffusion [Eq.~\eqref{eq:V-zero-ion-electron}] 
\begin{align}
    &\delta \mathcal{I}_{s}^{(0,1,0)} = 
    O_{\parallel}\left[ \mathcal{I}^{(0,1,1)}_s \right]
    \hat{b}^{\beta} \delta U_{\beta}
    -\frac{\delta b}{2 b}
    O_{\perp}\left[ \mathcal{I}^{(1,1,0)}_s \right]
    -
    \frac{q_s \mathrm{sgn}(k_z)}{m_s (i k_z)} \delta E_{\parallel} 
    \mathcal{H}^{(0,1,0)}_{s}
    \nonumber\\
    &+
    \frac{\mathrm{sgn}(k_z) \delta b}{2 b}
    \mathcal{H}^{(2,1,-1)}_{s} 
    = 0
    \,.
\end{align}
We eliminate $\hat{b}^{\beta} \delta U_{\beta}$ using the above equation to get
\begin{align}\label{eq:linearized-moments-final-expression}
    \delta \mathcal{I}^{(M,N,Q)}_s
    &=
    -\frac{q_s \delta E_{\parallel} \mathrm{sgn}(k_z)}{m_s (ik_z)}
    \bigg\{
    \frac{\mathcal{H}_s^{(0,1,0)} }{n_s}
    O_{\parallel}\left[ \mathcal{I}^{(M,N,Q+1)}_s \right]
    +
    \mathrm{sgn}^{N-1}(k_z)
    \mathcal{H}^{(M,N,Q)}_{s}
    \bigg\}
    \nonumber\\
    &+
    \frac{\delta b}{2b}
    \bigg\{
    \mathrm{sgn}^N(k_z)
    \mathcal{H}^{(M+2,N,Q-1)}_{s}
    +
    \frac{\mathrm{sgn}(k_z)
    }{n_s}
    \mathcal{H}^{(2,1,-1)}_{s}O_{\parallel}\left[ \mathcal{I}^{(M,N,Q+1)}_s \right]
    \nonumber\\
    &
    \hspace{5cm}
    -
    O_{\perp}\left[ \mathcal{I}^{(M+1,N,Q)}_s \right]
    \bigg\}
    \,.
\end{align}
This equation will be used as a starting point in Sec.~\ref{sec:Landau-Closure} to obtain the closures for the moments in Table~\ref{tab:moments-for-closure}. 
To evaluate the coefficients proportional to $\delta E_{\parallel}$ and $\delta b$, we analyze Eq.~\eqref{eq:Hfunc-definition}. In the non-relativistic limit, Eq.~\eqref{eq:Hfunc-definition} can be rewritten in terms of the plasma response function~\citep{1997PhPl....4.3974S}. The same simplification applies in relativity. For integer $Q>0$, one can show that (see Appendix~\ref{appendix:plasma-response-function})
\begin{align}\label{eq:H-zeta-expansion}
    \mathcal{H}^{(P,Q,R)}_s =
    -z_s \delta_{1,s}
    \sum_{k=0}^{Q} \zeta^k \, \mathcal{I}^{(-1)}_{(P,Q-k,R+1+k),s}
    -
    z_s \delta_{1,s} \mathcal{Z}^{(P,Q+R+1)}_s \zeta^{Q+1}
    \,,
\end{align}
where $\mathcal{Z}^{(P,R)}_s(\zeta)$ is a relativistic generalization of the plasma response function
\begin{align}\label{eq:Z-definition}
    \mathcal{Z}^{(P,R)}_s(\zeta) \equiv \left< \frac{v_{\perp}^P W^{R}}{\left(v_{\parallel} W_{U}^{-1} - \zeta\right) \Tilde{W}} \right>_s
    \,.
\end{align}
Given a value for $\zeta$, we evaluate all terms proportional to $\delta E_{\parallel}$ and $\delta b/b$ in Eq.~\eqref{eq:linearized-moments-final-expression} using Eqs.~\eqref{eq:H-zeta-expansion}, \eqref{eq:Oparallel-def}, and \eqref{eq:Operp-def}. Useful identities for numerically evaluating $\mathcal{Z}^{(P,R)}_s(\zeta)$, including analytical results in the high-temperature limit, are discussed in Appendix~\ref{appendix:plasma-response-function}.
\subsection{High-temperature and low-frequency limit}\label{sec:KT-heat-fluxes-linearized}
In this section, we provide analytical expressions for the first few moments of the linearized distribution function [Eq.~\eqref{eq:linearized-moments-final-expression}] in the high-temperature $z_{s}\ll 1$ and low frequency limit $\zeta \ll 1$.
Analytical expressions for the moments of the distribution function in this limit are given in Eq.~\eqref{eq:moments-distribution-function-analytical-appendix}.
The linearized moments from the KMHD equations in this limit can be obtained by using Eq.~\eqref{eq:linearized-moments-final-expression}
\begin{subequations}\label{eq:linearized-moments-analytical}
\begin{align}
    &\frac{\delta n_s}{n_s}
    =
    \frac{\delta b}{b } \left[-\tilde{\mu }-\frac{3}{4} i \pi  \zeta  (\tilde{\mu }+1)^{3/2}\right]
    \nonumber\\
    &+ \frac{q_s z_s \delta E_{\|}}{m_s k_z} \left[\frac{1}{2} \pi  \zeta  (\tilde{\mu }+1)^{3/2} \sqrt{\delta _2}-\frac{i \sqrt{\tilde{\mu }+1} \sqrt{\delta _2} \left(\sqrt{\tilde{\mu }}+(\tilde{\mu }+1) \tan ^{-1}\left(\sqrt{\tilde{\mu }}\right)\right)}{2 \sqrt{\tilde{\mu }}}\right]
    \,,\\
    &\frac{\delta p_{\parallel,s}}{p_{\parallel,s}}
    =
     \frac{\delta b}{b } \left[ 
     \frac{\left(-\tilde{\mu }^2+2 \tilde{\mu }+3\right) \tan ^{-1}\left(\sqrt{\tilde{\mu }}\right)-\sqrt{\tilde{\mu }} (\tilde{\mu }+3)}{2 (\tilde{\mu }+1) \tan ^{-1}\left(\sqrt{\tilde{\mu }}\right)-2 \sqrt{\tilde{\mu }}}
     \right]
     \nonumber\\
     &+
     \frac{q_s z_s \delta E_{\|}}{m_s k_z} \left[
     -\frac{2 i \tilde{\mu }^{3/2} \sqrt{\tilde{\mu }+1} \sqrt{\delta _2}}{3 (\tilde{\mu }+1) \tan ^{-1}\left(\sqrt{\tilde{\mu }}\right)-3 \sqrt{\tilde{\mu }}}\right]
    \,,\\
    &\frac{\delta p_{\perp,s}}{p_{\perp,s}}
    =
    \frac{\delta b}{b } \bigg[
    \frac{\left(-3 \tilde{\mu }^2+2 \tilde{\mu }-3\right) \tan ^{-1}\left(\sqrt{\tilde{\mu }}\right)-3 (\tilde{\mu }-1) \sqrt{\tilde{\mu }}}{2 \left(\sqrt{\tilde{\mu }}+(\tilde{\mu }-1) \tan ^{-1}\left(\sqrt{\tilde{\mu }}\right)\right)}
    \nonumber\\
    &\hspace{3cm}
    -\frac{2 i \pi  \zeta  \tilde{\mu }^{3/2} (\tilde{\mu }+1)}{\sqrt{\tilde{\mu }}+(\tilde{\mu }-1) \tan ^{-1}\left(\sqrt{\tilde{\mu }}\right)}
     \bigg]
     \nonumber\\
     &+
     \frac{q_s z_s \delta E_{\|}}{m_s k_z} \left[
     \frac{\pi  \zeta  \tilde{\mu }^{3/2} (\tilde{\mu }+1) \sqrt{\delta _2}}{\sqrt{\tilde{\mu }}+(\tilde{\mu }-1) \tan ^{-1}\left(\sqrt{\tilde{\mu }}\right)}-\frac{4 i \tilde{\mu }^{3/2} \sqrt{\tilde{\mu }+1} \sqrt{\delta _2}}{3 \left(\sqrt{\tilde{\mu }}+(\tilde{\mu }-1) \tan ^{-1}\left(\sqrt{\tilde{\mu }}\right)\right)}
     \right]
    \,,\\
    &\frac{\delta \mathcal{Q}_{\parallel,s}}{\mathrm{sgn}(k_z)}
    =
    \frac{\delta b}{b }\frac{\pi  \alpha  \zeta  m_s}{z_s^4} \left[ 
    \frac{6 \left(5 \tilde{\mu }^2+14 \tilde{\mu }+9\right) \tan ^{-1}\left(\sqrt{\tilde{\mu }}\right)}{\tilde{\mu }^{5/2} \delta _2{}^2}-\frac{6 (11 \tilde{\mu }+9)}{\tilde{\mu }^2 \delta _2{}^2}
     \right]
     \nonumber\\
     &+
     \frac{\pi  \alpha  \zeta  q_s z_s E_{\|}}{k_z}  \bigg[
     -\frac{12 i \sqrt{\tilde{\mu }+1} \tan ^{-1}\left(\sqrt{\tilde{\mu }}\right)}{\tilde{\mu }^{3/2} \delta _2{}^{3/2} z_s^4}+\frac{18 i (\tilde{\mu }+1)^{5/2} \tan ^{-1}\left(\sqrt{\tilde{\mu }}\right)^2}{\tilde{\mu }^3 \delta _2{}^{3/2} z_s^4}
     \nonumber\\
     &
     \hspace{3cm}
     -\frac{2 i \left(4 \tilde{\mu }^2+15 \tilde{\mu }+9\right)}{\tilde{\mu }^2 \sqrt{\tilde{\mu }+1} \delta _2{}^{3/2} z_s^4}
     \bigg]
    \,,\\
    &\frac{\delta \mathcal{Q}_{\perp,s}}{\mathrm{sgn}(k_z)}
    =
    \frac{\delta b}{b }\frac{\pi  \alpha  \zeta  m_s}{z_s^4} \left[ 
    -\frac{3 (\tilde{\mu }-9) (\tilde{\mu }+1)}{\tilde{\mu }^2 \delta _2{}^2}-\frac{3 \left(\tilde{\mu }^2+2 \tilde{\mu }+9\right) (\tilde{\mu }+1) \tan ^{-1}\left(\sqrt{\tilde{\mu }}\right)}{\tilde{\mu }^{5/2} \delta _2{}^2}
     \right]
     \nonumber\\
     &+
     \frac{\pi  \alpha  \zeta  q_s z_s E_{\|}}{k_z}  
     \bigg[
     \frac{6 i (\tilde{\mu }+1)^{3/2} \tan ^{-1}\left(\sqrt{\tilde{\mu }}\right)}{\tilde{\mu }^{3/2} \delta _2{}^{3/2} z_s^4}+\frac{3 i (\tilde{\mu }-3) (\tilde{\mu }+1)^{5/2} \tan ^{-1}\left(\sqrt{\tilde{\mu }}\right)^2}{\tilde{\mu }^3 \delta _2{}^{3/2} z_s^4}
     \nonumber\\
     &+\frac{i \left(-8 \tilde{\mu }^2+3 \tilde{\mu }+9\right) \sqrt{\tilde{\mu }+1}}{\tilde{\mu }^2 \delta _2{}^{3/2} z_s^4}
     \bigg]
    \,,\\
    \label{eq:Qs-closure}
    &\delta \mathcal{Q}_{s}
    =
    \delta \mathcal{Q}_{\parallel,s}
    +
    2\delta \mathcal{Q}_{\perp,s}
    \,.
\end{align}
\end{subequations}
The quantities $\tilde{\mu}$ and $\tilde{\Delta}$ are defined in Eq.~\eqref{eq:mutilde-Delta-tilde}.
The expressions above will be analyzed below to provide a Landau-fluid closure for the heat fluxes; observe that we have already obtained a closure for $\mathcal{Q}_s$ in Eq.~\eqref{eq:Qs-closure}.
\section{Landau fluid closure}\label{sec:Landau-Closure}
In this section, we derive the closure relationship for the moments of the distribution function listed in Table~\ref{tab:moments-for-closure}.
We first outline our closure strategy in Section~\ref{sec:closure-strategy} and present an analytical closure relationship in Sec.~\ref {sec:closure-analytical}.
While our approach can be used to obtain the closure for different relativistic equations of state, analytical closures can only be obtained either in the non-relativistic limit $(m \gg T)$~\citep{1997PhPl....4.3974S} or in the ultra-relativistic limit $(m\ll T)$.
This is not a serious problem, since we want to apply our models to an M87-like system, where the ultra-relativistic equation state is a reasonable assumption for electrons~\citep{EHT-2021,EventHorizonTelescope:2019pcy}.
For the ions, one can use the non-relativistic heat fluxes from~\citet{1997PhPl....4.3974S}.

In Sec.~\ref {sec:compare-kinetic-and-fluid}, we compare the linearized evolution of the energy and density response from KMHD with the MHD, CGL, and Landau fluid closures.
In Sec.~\ref {sec:collisions}, we discuss how one includes the effect of collisions in the heat fluxes phenomenologically.
We conclude this section with a discussion of how our model differs from other approaches in the literature. 
We first compare our model with the collisional models considered in~\citep{Chandra_2015,Most_2021} in Sec.~\ref{sec:compare-models-coll}.
In Sec.~\ref{sec:compare-new-literature}, we compare our method to independent studies--which we became aware of during the final stages of submission--~\citet{wierzchucka2026doubleadiabaticequationsstaterelativistic,ley2026doubleadiabaticequationsrelativisticregime} which derive the CGL equations for relativistic equations of state assuming spatial homogeneity and non-relativistic flows. 
\subsection{Closure strategy}\label{sec:closure-strategy}
If collisions dominate a kinetic system, then the distribution function evolves toward a local equilibrium characterized by macroscopic properties such as fluid velocity, energy density, and pressure. For small departures from local equilibrium in a collisional system, there are well-developed strategies such as the Chapman-Enskog approximation~\citep{1991mtnu.book.....C} or Grad's moment approximation~\citep{Grad:1949zza} that allow one to systematically understand the evolution of the moments of the distribution functions and calculate non-equilibrium contributions such as viscous stresses in the system. For a general discussion of these techniques in the relativistic context, see~\citet{2021mfrf.book.....D,Denicol_2018,Denicol_2012}; for numerical simulations of black hole accretion flow in the collisional Braginskii limit, see~\citet{Chandra_2015,Foucart_2015,Foucart_2017,dhruv2025electromagneticobservablesweaklycollisional}; for discussions of collisional closure in multi-species magnetized plasma, see~\citep{Most_2021,Most_2022}.

The plasmas studied in this paper are weakly collisional [Eq.~\eqref{eq:ordering-check}].
To obtain a closure for the heat fluxes for these systems, we adopt the Landau fluid closure introduced in~\citep{PhysRevLett.64.3019,1997PhPl....4.3974S,Goswami_2005}
\begin{itemize}
    \item First, define a set of master variables. In this paper, these variables will be $(\mathcal{E}_s/n_s, p_{\perp,s} - p_{\parallel,s}, b, n_s )$.
    \item Choose a moment of the distribution function that requires a closure. For illustration, suppose that this is $\mathcal{Q}_{\parallel,s}$.
    \item Obtain the linearized expression for $\delta \mathcal{Q}_{\parallel,s}$ using Eq.~\eqref{eq:linearized-moments-final-expression} in the low-frequency limit.
    Schematically, this is of the form
    \begin{align}\label{eq:lin-schematic}
        \delta \mathcal{Q}_{\parallel,s} = \left[s_1 + s_2 \zeta \right] \delta E_{\parallel}
        +
        \left[s_1 + s_2 \zeta\right] \delta b
        \,,
    \end{align}
    where the coefficients $s_{i}$ depend on the equilibrium distribution function [Eq.~\eqref{eq:anisotropic-distribution-function-relativity}].
    \item Consider a closure of the form
    \begin{align}\label{eq:ansatz-for-closure}
        \delta \mathcal{Q}_{\parallel,s} = 
        i \mathrm{sgn}(k_z) 
        \bigg[r_1 n_s \delta\left(\frac{\mathcal{E}_s}{n_s}\right) 
        + 
        r_2 
        \delta\left(p_{\perp,s} - p_{\parallel,s}\right) 
        + 
        r_3 \frac{\delta b}{b} \mathcal{E}_s
        + 
        r_4  
        \mathcal{E}_s
        \frac{\delta n_s}{n_s}
        \bigg]
        \,.
    \end{align}
    Equate Eq.~\eqref{eq:lin-schematic} and \eqref{eq:ansatz-for-closure} and determine $r_{i}$.  
\end{itemize}
In the non-relativistic limit, one can interpret this strategy as a procedure to build Pad\'{e} approximations to the plasma response function, see~\citet{1997PhPl....4.3974S,Hunana_2019,Goswami_2005}.
This interpretation works at the linear level, mainly due to the choice of a bi-Maxwellian distribution function which allows one rewrite all the linearized moments in terms of the plasma response function, see Eqs.~(36)-(38) of~\citet{1997PhPl....4.3974S}. In relativity, we interpret this approach as a way to construct an effective field theory for heat fluxes that closely reproduces the kinetic dispersion relation and captures Landau damping.
\subsection{Analytical closure in the ultra-relativistic limit}\label{sec:closure-analytical}
If we implement the Landau fluid closure strategy outlined above to Eq.~\eqref{eq:linearized-moments-analytical}, we see that the heat fluxes satisfy
\begin{subequations}\label{eq:closure-analytical-heat-fluxes}
\begin{align}
    \delta \mathcal{Q}_{s}
    &=
    \delta \mathcal{Q}_{\parallel,s}
    +
    2\delta \mathcal{Q}_{\perp,s} \,,\\
    \delta \mathcal{Q}_{\parallel,s} &= 
    i \mathrm{sgn}(k_z)
    \bigg[ 
    n_s
    \left(
    -\frac{8}{5\pi}
    +
    f_1
    \right)
    \delta\left(\frac{\mathcal{E}_s}{n_s}\right)  
    +
    \left(
    \frac{28}{15 \pi}
    +
    f_2
    \right)
    \delta\left(p_{\perp,s}-p_{\parallel,s}\right)
    \nonumber\\
    &+
    f_3 \frac{\delta b}{b} \mathcal{E}_s
    +
    f_4 \mathcal{E}_s \frac{\delta n_s}{n_s}
    \bigg]
    \,,\\
    \delta \mathcal{Q}_{\perp,s} &= 
    i \mathrm{sgn}(k_z)
    \bigg[ 
    n_s
    \left(
    -\frac{8}{15\pi}
    +
    g_1
    \right)
    \delta\left(\frac{\mathcal{E}_s}{n_s}\right)  
    +
    \left(
    -\frac{4}{15 \pi}
    +
    g_2
    \right)
    \delta\left(p_{\perp,s}-p_{\parallel,s}\right)
    \nonumber\\
    &+
    g_3 \frac{\delta b}{b} \mathcal{E}_s
    +
    g_4 \mathcal{E}_s \frac{\delta n_s}{n_s}
    \bigg]
    \,.
\end{align}
\end{subequations}
The coefficients $f_{1,2,3,4}$ and $g_{1,2,3,4}$ are complicated functions of $\tilde{\mu}$ and vanish when the initial anisotropy vanishes. We list these coefficients in the supplementary \texttt{Mathematica} file.
One could repeat the Landau fluid procedure to obtain an accurate closure for $r_{\perp,\perp}^{(-2)} , r_{\parallel,\parallel}^{(-2)}$ and $r_{\parallel,\perp}^{(-2)}$.
However, we find that it is easier to use the equilibrium relationship for these functions.
In the high temperature limit $(z\to 0 \implies m \ll T)$, these are given by
\begin{align}\label{eq:closure-higher-order-moments-analytical}
    r_{\perp,\perp,s}^{(-2)} &= 
    \mathcal{E}_s
    \left[
    \frac{2}{15} + h_1
    \right]
    \,, \quad 
    r_{\parallel,\parallel,s}^{(-2)} = 
    \mathcal{E}_s
    \left[
    \frac{1}{5}
    + h_2
    \right]
    \,, \quad 
    r_{\parallel,\perp,s}^{(-2)} = 
    \mathcal{E}_s
    \left[
    \frac{1}{15}
    + h_3
    \right]
    \,,
\end{align}
where $h_{1,2,3}$ are complicated functions of $\tilde{\mu}$ that vanish when the pressure anisotropy is zero. These coefficients are provided in the supplementary \texttt{Mathematica} file.

Equations~\eqref{eq:closure-analytical-heat-fluxes}, \eqref{eq:closure-higher-order-moments-analytical} and \eqref{eq:V-zero-ion-electron} provide the closure relationship for all the moments listed in Table~\ref{tab:moments-for-closure} in the ultra-relativistic limit for small pressure anisotropy. These Landau fluid closures can be used to evolve the fluid equations [Eq.~\eqref{eq:all-species-equations-3+1}] instead of the kinetic equations to capture weakly collisional effects; see~\citet{Sharma_2006,2007ApJ...667..714S,Squire2023} for numerical simulations in the non-relativistic limit.

Before we proceed, let us briefly compare the heat flux closure in the ultra-relativistic limit [Eq.~\eqref{eq:closure-analytical-heat-fluxes}] with the non-relativistic limit from Eqs. (48) and (49) of~\citet{1997PhPl....4.3974S} 
\begin{subequations}\label{eq:closure-non-relativistic}
\begin{align}
    &\delta \mathcal{Q}_{\parallel}^{\mathrm{non-rel}} = - \sqrt{\frac{8}{\pi }} \frac{n_s i \mathrm{sgn}(k_z) }{\sqrt{z_{\parallel,s}} }
    \delta \left(\frac{ p_{\parallel,s}}{n_s}\right)
    \,,
    \quad
    z_{\parallel} \equiv n_s/p_{\parallel,s}
    \,,
    \\
    &\delta \mathcal{Q}_{\perp}^{\mathrm{non-rel}} = \
    - \sqrt{\frac{2}{\pi }} \frac{n_s i \mathrm{sgn}(k_z) }{ \sqrt{z_{\parallel,s}}  }
    \bigg[
    \delta \left(\frac{ p_{\perp}}{n_s}\right)
    +
    \frac{m_s (p_{\perp,s} - p_{\parallel,s})}{p_{\parallel,s}} \frac{\delta b}{b z_{\perp,s}}
    \bigg]
    \,,
    z_{\perp} \equiv n_s/p_{\perp,s}
    \,.
\end{align}
\end{subequations}
Observe that the non-relativistic parallel (perpendicular) heat flux is driven by the gradients in the parallel (perpendicular) pressure.
However, in the relativistic heat fluxes, the anisotropic distribution is spheroidal [Eq.~\eqref{eq:anisotropic-distribution-function-relativity}] and both parallel and perpendicular pressure gradients contribute. 

In relativistic flows applicable to SMBHs, the electrons are often ultra-relativistic,
and one can use the ultra-relativistic closure relation in Eq.~\eqref{eq:closure-higher-order-moments-analytical} to evolve their heat fluxes.
For the ions, which are typically non-relativistic or only mildly relativistic, one can instead use the covariant form of the non-relativistic closure in
Eq.~\eqref{eq:closure-non-relativistic}. More generally, the same matching procedure used in this section can in principle be repeated at finite temperature~\citep{Denicol_2014}. In that case, the relevant kinetic response functions and moment integrals depend on the dimensionless temperature $z\sim m/T$, and the closure coefficients need not take the simple analytic form found in the ultra-relativistic limit. They may instead be obtained by constructing asymptotic expansions about the non-relativistic and ultra-relativistic limits~\citep{Denicol_2014}, or by evaluating the finite-temperature response functions numerically. Hence, the ultra-relativistic and non-relativistic closures should be viewed as limiting cases of a more general finite-temperature relativistic Landau-fluid closure.
\begin{figure*}
    \centering
    \includegraphics[width=1\linewidth]{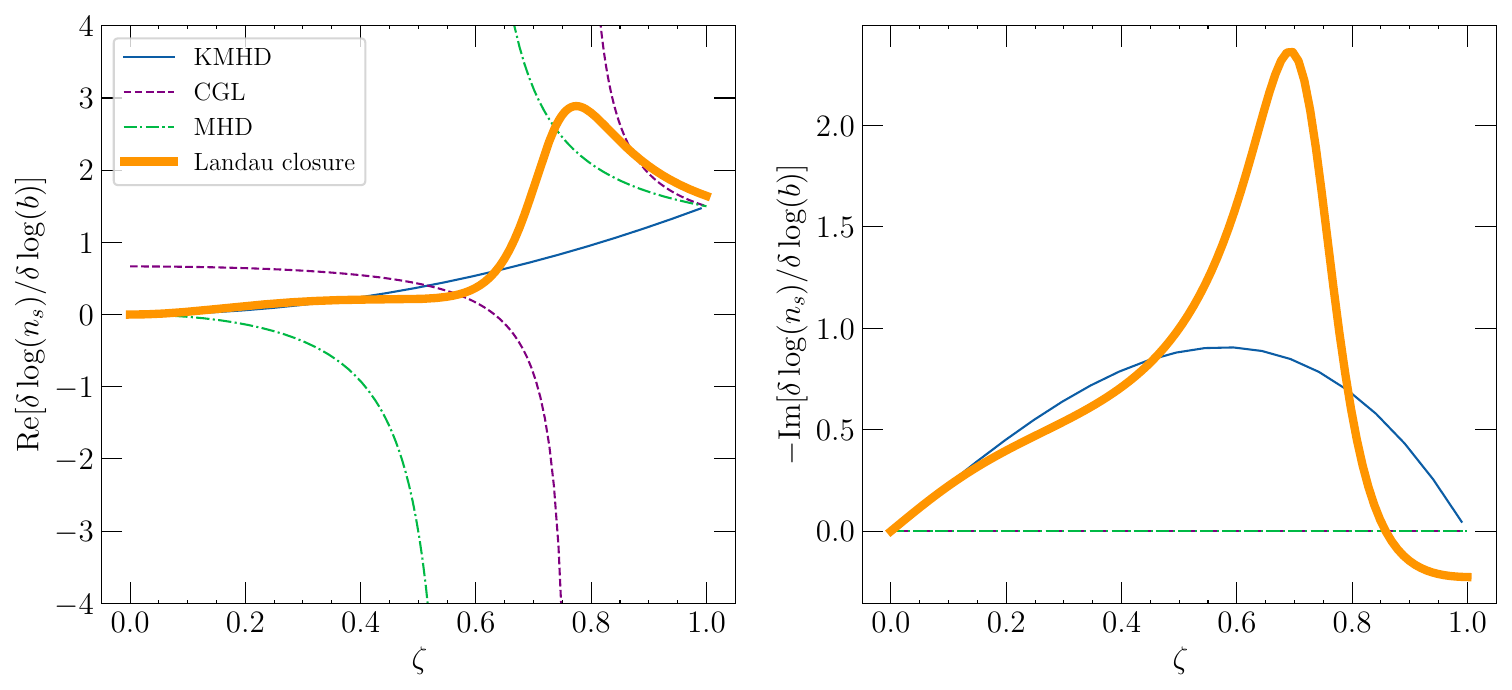}
    \includegraphics[width=1\linewidth]{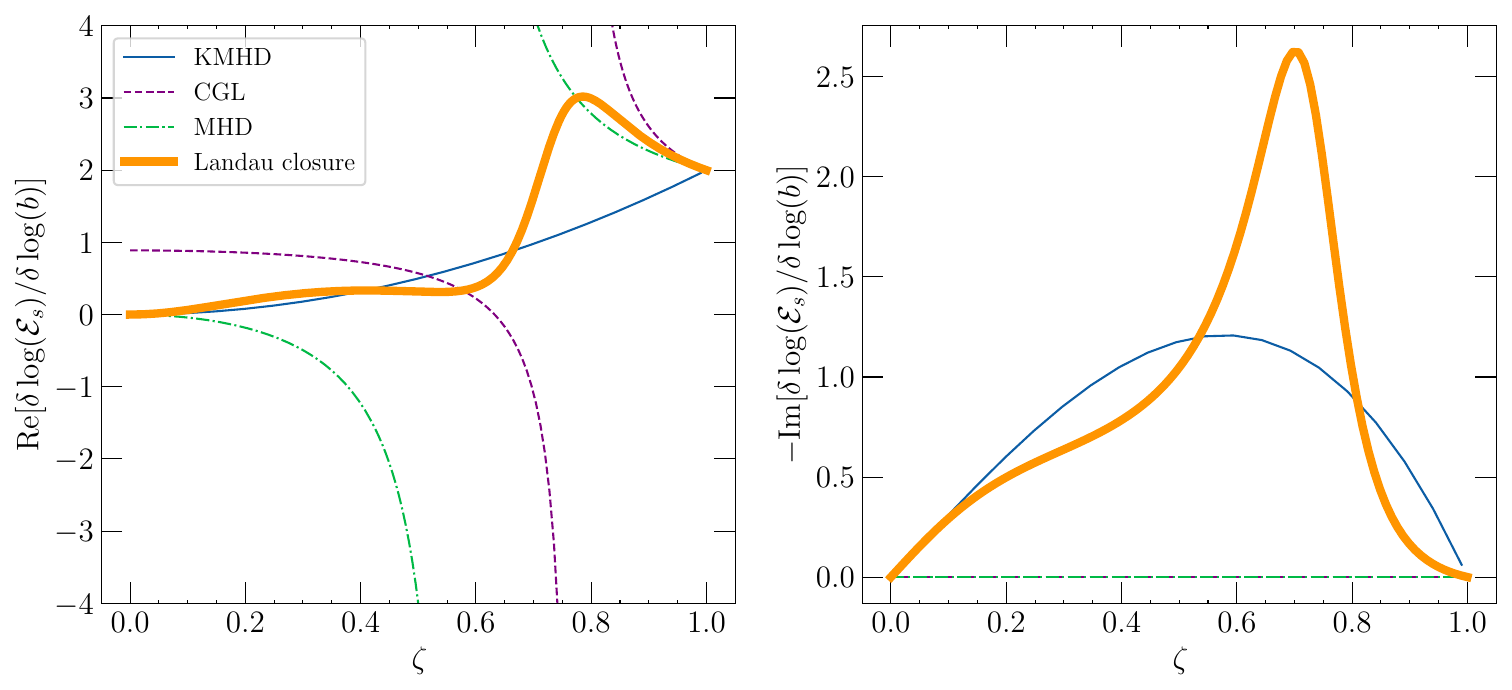}
    \caption{Comparison between the linear number density (top panel) and energy response (bottom panel) between the KMHD (solid blue line), CGL (dashed purple line), MHD (dash dotted green line) and the Landau fluid closure (bold orange line). as a function of the driving speed $\zeta = \omega/|k_z|$. 
    The left and right columns compares the real and imaginary parts of the response. Observe from the right column that the Landau fluid closure qualitatively models the Landau damping predicted by KMHD equations. This damping is not present in the CGL or the MHD models. }
    \label{fig:response_comparison}
\end{figure*}
\subsection{Comparison between kinetic and fluid response}\label{sec:compare-kinetic-and-fluid}
In this section, we compare the density and energy density responses obtained from the Landau fluid closure and with the full KMHD calculation.
For simplicity, as in the non-relativistic calculation~\citep{1997PhPl....4.3974S}, we choose to compare the responses when the initial distribution function is isotropic.
In this case, all the $\Delta p$ dependent terms drop out from the Landau fluid closure relations.
The linearized Landau fluid system can be obtained from Eq.~\eqref{eq:fluid-eom-species-s} to obtain a response equation similar to the KMHD result, see, Eq.~\eqref{eq:linearized-moments-analytical}.
The exact results for the fluid response from the MHD, CGL and Landau fluid closures are listed in Appendix~\ref{appendix:linearized-fluid-and-MHD}.
We also include a comparison between the fluid response and the kinetic response in case of an initially anisotropic distribution in Appendix~\ref{appendix:initially-anisotropic-distribution}.

The top and bottom panel of Fig.~\ref{fig:response_comparison}, compares the KMHD response results with the Landau fluid closure for the number density response and the energy density, respectively as a function of the driving speed $\zeta = \omega/|k_z|$. 
The left and right panel of each row plots the real and imaginary part of the response functions.
Observe from the left column that the real part of the response functions for the number density and the energy density match the KMHD response function in the low frequency regime.
The MHD (green dash dotted line) and CGL (purple dashed line) models have resonances at sound speeds $\zeta = \sqrt{1/3}$ and $\zeta = \sqrt{3/5}$, respectively.
This resonant behavior is absent in the KMHD model (blue solid line) and the Landau fluid closure (bold orange line). 
From the right column of the figure, we see that the Landau fluid closure approximately captures the Landau damping predicted by the KMHD equations, a feature that this not present in either the CGL or the MHD equations. However, the Landau fluid closure over damps the system at $\zeta \sim 0.65$.
This behavior is also observed in the non-relativistic theory; compare Fig.~\ref{fig:response_comparison} with Fig. 3 of~\cite{1997PhPl....4.3974S}.
For initially anisotropic distribution function, we observe results qualitatively similar to Fig.~\ref{fig:response_comparison}, see Fig.~\ref{fig:response_comparison_aniso} in Appendix~\ref{appendix:initially-anisotropic-distribution}.

The results from Fig.~\ref{fig:response_comparison} show that the Landau fluid closure obtained in the previous section reproduces the dissipative KMHD response in the low-frequency limit and qualitatively captures the Landau damping predicted by the kinetic equations.
At intermediate and high phase speeds, however, quantitative differences between the Landau fluid closure and the exact KMHD response remain.
The agreement is expected to improve as more moments are retained in the closure relationship.
While this has been demonstrated in the non-relativistic limit (see Fig.~3 of~\citet{1997PhPl....4.3974S}), here we do not explore such higher-order closures, as it is unclear at present how to incorporate them consistently into a nonlinear relativistic simulation.
Moreover, in the absence of a precise order-counting scheme, improved convergence in linear theory does not necessarily imply improved behavior in a nonlinear simulation.
We therefore choose a simple closure that captures some of the essential kinetic physics while keeping the resulting fluid model tractable.
\subsection{Collisional operator and phenomenological closure}\label{sec:collisions}
Let us now include the effects of collisions using a simplified model. We ignore inter-species collisions and use a simple linear operator operator introduced in~\citet{Rocha-Denicol-Jorge-RTA} to model the collisions
\begin{align}
    \mathcal{C}_s^{\mathrm{RTA}}[f_s]
    &=
    -\frac{W_U}{\tau_{R,s}}
    \bigg[
    \boldsymbol{\delta} f_s
    -
    \feq{s}
    \frac{\left<W_U \delta f\right>}{\left<W_U \feq{s} \right>}
    -
    P_{1} 
    \frac{\feq{s} \left<W_U P_{1} \boldsymbol{\delta} f_s \right>}{\left<W_U P_{1} P_{1} \feq{s} \right>}
    \nonumber\\
    &-
    k^{<\mu>}
    \frac{ \feq{s} \left<W_U k_{<\mu>} \boldsymbol{\delta} f_s \right>}{(1/3) \left<W_U k_{<\nu>} k^{<\nu>} \feq{s} \right>
    }
    \bigg]
    \,,
\end{align}
where $\feq{s}$ is the Maxwell-Juttner distribution function
\begin{align}
    \feq{s} = \alpha_{s,\mathrm{eq}} \exp(- z_{s,\mathrm{eq}} \sqrt{1 + v_{\parallel}^2 + v_{\perp}^2})\,,
\end{align}
$\boldsymbol{\delta} f_s = f_s - \feq{s}$, $\tau_{R,s}$ is a relaxation time, and, 
\begin{align}
    P_1 &\equiv 
    1-\frac{\left<W_U \feq{s} \right>}{\left<W_U^2 \feq{s} \right>} W_U
    \,.
\end{align}
Suppose that the plasma is ultra relativistic and choose the equilibrium function so that it satisfies
\begin{align}
    \left<W_U \feq{s} \right> = \left<W_U f_s\right>, \left<W_U^2 \feq{s} \right> = \left<W_U^2 f_s\right> \implies \alpha_{s,\mathrm{eq}} = \frac{\alpha_s}{\sqrt{\delta_1} \delta_2} \,,
    z_{s,\mathrm{eq}} = z
    \,.
\end{align}
These assumptions simplify the collisional operator to
\begin{align}
    \mathcal{C}_s^{\mathrm{RTA}}[f_s]
    &=
    -\frac{W_U}{\tau_{R,s}}
    \bigg[
    \boldsymbol{\delta} f_s
    -
    k^{<\mu>}
    \frac{ \feq{s} \left<W_U k_{<\mu>} \boldsymbol{\delta} f_s \right>}{(1/3) \left<W_U k_{<\nu>} k^{<\nu>} \feq{s} \right>
    }
    \bigg]
    \,.
\end{align}
For the gyro-averaged kinetic equation, we need the gyro-averaged collisional operator which takes the form
\begin{align}\label{eq:linearized-collisional-operator-final}
    \mathcal{C}_s[f_s]
    =
    \frac{1}{2\pi}
    \int d\vartheta\, \mathcal{C}_s^{\mathrm{RTA}}[f_s]
    =
    -\frac{W_U}{\tau_{R,s}}
    \bigg[
    \boldsymbol{\delta} f_s
    -
    v_{\parallel}
    \frac{ \feq{s} \mathcal{Q}_s}{p_s
    }
    \bigg]
    \,.
\end{align}
We can use the above expression to simply the $O_{c}\left[ \mathcal{I}^{(0,2,-1)}_s \right]$ and $O_{c}\left[ \mathcal{I}^{(2,0,-1)}_s \right]$ appearing in Eqs.~\eqref{eq:pparallel-Eqn-all-species} and \eqref{eq:pperp-Eqn-all-species}
\begin{subequations}
\begin{align}
    O_{c}\left[ \mathcal{I}^{(0,2,-1)}_s \right] &= -\frac{2}{3 \tau_{R,s} m_s} \left(p_{\parallel,s} - p_{\perp,s}\right) \,,\\
    O_{c}\left[ \mathcal{I}^{(2,0,-1)}_s \right] &= 
    \frac{2}{3 \tau_{R,s} m_s} \left(p_{\parallel,s} - p_{\perp,s}\right)
    \,.
\end{align}
\end{subequations}
These equations are exactly equivalent to their non-relativistic counter parts, see, Eqs. (16) and (17) of~\citet{1997PhPl....4.3974S}.

It is possible to include the collisional operator in the kinetic equation to obtain a collisional closure. However, performing this exactly is not feasible because the collisional operator introduces poles in the Green's function that are hard to treat. In the non-relativistic limit,~\citet{1997PhPl....4.3974S} followed an alternate approach: first deriving a collisionless closure for higher order moments of the distribution function, then taking the low-frequency limit with collisions included to obtain the closure for the heat fluxes. This procedure can also be carried out in relativity, but one typically needs to close more moments than in the non-relativistic limit. The result of this calculation will be a closure of the form
\begin{subequations}\label{eq:heat-flux-closure-with-collisions}
\begin{align}
    \delta \mathcal{Q}_{\parallel,s} &= 
    i \frac{|k_z|}{k_z + \tau_{R,s}^{-1} \boldsymbol{a}_{\parallel,s}}
    \bigg[ 
    n_s
    \left(
    -\frac{8}{5\pi}
    +
    f_1
    \right)
    \delta\left(\frac{\mathcal{E}_s}{n_s}\right)  
    +
    \left(
    \frac{28}{15 \pi}
    +
    f_2
    \right)
    \delta\left(p_{\perp,s}-p_{\parallel,s}\right)
    \nonumber\\
    &+
    f_3 \frac{\delta b}{b} \mathcal{E}_s
    +
    f_4 \mathcal{E}_s \frac{\delta n_s}{n_s}
    \bigg]
    \,,\\
    \delta \mathcal{Q}_{\perp,s} &= 
    i \frac{|k_z|}{k_z + \tau_{R,s}^{-1} \boldsymbol{a}_{\perp,s}}
    \bigg[ 
    n_s
    \left(
    -\frac{8}{15\pi}
    +
    g_1
    \right)
    \delta\left(\frac{\mathcal{E}_s}{n_s}\right)  
    +
    \left(
    -\frac{4}{15 \pi}
    +
    g_2
    \right)
    \delta\left(p_{\perp,s}-p_{\parallel,s}\right)
    \nonumber\\
    &+
    g_3 \frac{\delta b}{b} \mathcal{E}_s
    +
    g_4 \mathcal{E}_s \frac{\delta n_s}{n_s}
    \bigg]
    \,,
\end{align}
\end{subequations}
where $\boldsymbol{a}_{\parallel,s}$ and $\boldsymbol{a}_{\perp,s}$ are $\mathcal{O}(1)$ coefficients. Notice that when $\tau_{R,s} \to \infty$ the above equation reduces to the collisionless closure Eq.~\eqref{eq:closure-analytical-heat-fluxes}.
In a numerical implementation, one can simply treat $\boldsymbol{a}_{\parallel,s}$ and $\boldsymbol{a}_{\perp,s}$ as phenomenological parameters that decrease the pressure anisotropy~\citep{Sharma_2006,Squire2023}.

Before proceeding, it is useful to comment on the locality, stability, and
causality properties of the closure in Eq.~\eqref{eq:heat-flux-closure-with-collisions}.
The closure was derived from the linearized equations in Fourier space. In the
collisionless limit, the appearance of factors such as \(\mathrm{sgn}(k_z)\) implies that,
when transformed back to physical space, the closure is nonlocal along the magnetic
field. This is the familiar Hilbert-transform-like nonlocality of Landau-fluid closures
\citep{PhysRevLett.64.3019,1997PhPl....4.3974S}.

This nonlocality should not be interpreted as a fundamental inconsistency of the
underlying relativistic kinetic theory. Rather, it reflects the fact that the closure is a
low-frequency representation of collisionless phase mixing. In practical fluid
simulations, one often replaces the nonlocal operator by a local phenomenological
approximation involving a characteristic parallel wavenumber \(k_z\), see~\citet{Sharma_2006}.

Even after such a replacement, the resulting equations should not be viewed as a
manifestly causal, strictly hyperbolic relativistic fluid theory. This is expected, since
the closure was constructed in a low-frequency approximation rather than from a
relaxation-type causal theory. A practical way to handle this limitation is to use an
order-reduction scheme: time derivatives appearing in the heat-flux sector are replaced
using the lower-order CGL equations, thereby expressing the heat-flux corrections in
terms of spatial derivatives of the leading-order variables. Such an order-reduced
system can be evolved stably provided the heat-flux corrections remain perturbative,
i.e. small compared with the leading CGL terms.
\subsection{Comparisons to models in literature}\label{sec:compare-models}
Before we compare our model to other relativistic flow models, let us briefly place it in the context of non-relativistic models. Historically, the description of magnetized, weakly collisional plasmas relied on the Braginskii fluid equations~\citep{1958JETP....6..358B}, which requires a viscous closure for the heat-fluxes. However, in many astrophysical environments—such as the solar wind, the intracluster medium, and accretion flows—the mean free path of particles exceeds the scale lengths of interest, requiring collisionless models~\citep{PhysRevLett.64.3019,1997PhPl....4.3974S,2009ApJS..182..310S}.

In the context of accretion flows around SMBHs, several studies have used collisionless Landau-fluid models to study the impact of pressure anisotropy in a local shearing-box simulation~\citep{Sharma_2003,Sharma_2006}. The model developed in this paper will help generalize these studies to black hole accretion flows to understand the impact of pressure anisotropy on relativistic jets and near-horizon flow dynamics. Beyond basic transport, the model studied here should allow the investigation of plasma dynamics absent in standard ideal GRMHD, such as the firehose and mirror instabilities~\citep{Kunz_2014} and magnetic immutability~\citep{Squire_2019,Squire2023,majeski2024selforganizationcollisionlesshighbetaturbulence}. 

\subsubsection{Comparison to EMHD}\label{sec:compare-models-coll}
In this section, we briefly compare the Landau fluid model presented here with the EMHD models developed by \citet{Chandra_2015,Most_2021,Most_2022}. 
Both approaches yield an identical structural form for the stress-energy tensor, where non-equilibrium contributions are driven by pressure anisotropy and magnetic-field-aligned heat fluxes [Eq.~\eqref{eq:gyrotropic-density-stress-components}].
However, the underlying closure physics is fundamentally different.
The original model of~\citet{Chandra_2015} was a phenomenological model for a plasma in which micro-instabilities have already relaxed plasma to a local equilibrium state. Our model, however, does not make this assumption and instead tries to capture general pressure anisotropies. 

More specifically, the form of the heat flux [Eq.~\eqref{eq:closure-analytical-heat-fluxes}] and the pressure evolution equations [Eqs.~\eqref{eq:pparallel-Eqn-all-species} and \eqref{eq:pperp-Eqn-all-species}] differ significantly between the two models. In the EMHD approach, the system is collision-dominated, and the distribution function is assumed to remain close to a local Maxwell-Juttner equilibrium. This assumption restricts the magnitude of the pressure anisotropy and heat flux, forcing these out-of-equilibrium corrections to relax toward a Braginskii-like form (see Eqs. (45) and (50) of \citet{Chandra_2015}):
\begin{subequations}
\begin{align}
    &\mathcal{Q}_{\mathrm{EMHD}} \sim - \rho \chi \hat{b}^{\mu} \left[ \nabla_{\mu} \Theta + \Theta a_{\mu} \right] \,,\\
    &\left(p_{\perp} - p_{\parallel}\right)_{\mathrm{EMHD}} \sim 3 \rho \nu \left(\hat{b}^{\mu} \hat{b}^{\nu} \nabla_{\mu} U_{\nu} - \frac{1}{3} \nabla_{\mu} U^{\mu} \right)
    \,,
\end{align}
\end{subequations}
where $\chi$, $\nu$ and $\Theta$ are the thermal transport coefficient, the viscous transport coefficients and the temperature. Unlike the EMHD model, the initial distribution function in the Landau fluid is assumed to be anisotropic and non-Maxwellian. This allows both parallel and perpendicular pressure variables to evolve dynamically. Moreover, the sources of dissipation in the two models differ. The EMHD models dissipation through local collisions, whereas in the Landau fluid model, dissipation arises from kinetic Landau damping and phase-mixing. 
\subsubsection{Comparison to recent weakly collisional models}\label{sec:compare-new-literature}
During the final stages of preparing this manuscript, two recent studies by \citet{wierzchucka2026doubleadiabaticequationsstaterelativistic} and \citet{ley2026doubleadiabaticequationsrelativisticregime} were submitted.
These studies generalize the CGL equations to relativistic \textit{distribution functions} with the assumptions of spatial homogeneity and non-relativistic bulk flow.
Here, we briefly describe how to obtain the results from these papers using our more general approach. Consider a collisionless, homogeneous plasma with non-zero parallel electric, non-relativistic bulk flow ($W_U \sim 1$) in Minkowski space and assume that $\partial_t f \gg \partial_{x,y,z} f$.
Using these assumptions Eq.~\eqref{eq:gyroaveraged-kinetic-equation-summary} reduces to Eq. (2.2) of~\citet{ley2026doubleadiabaticequationsrelativisticregime}
\begin{align}
    \partial_t f_s + 
    \frac{v_{\perp} \partial_t b }{2 b} 
    \frac{\partial f_{s}}{\partial v_{\perp}}
    \!+\!
    v_{\parallel}
    \frac{\partial f_{s}}{\partial v_{\parallel}}
    \left(\frac{\partial_t n_s}{n_s} - \frac{\partial_t b}{b} \right)
    =
    0\,.
\end{align}
The above equation can be solved using the method of lines to obtain Eqs. (2.5), (2.11) and (2.12) of~\citet{ley2026doubleadiabaticequationsrelativisticregime}.
Once the evolution of the distribution function is obtained, \citet{ley2026doubleadiabaticequationsrelativisticregime} derive the evolution of the parallel and perpendicular pressures for different initially isotropic distribution functions such as the Maxwell-Boltzmann distribution and the Maxwell-Juttner distribution.
The results of~\citet{wierzchucka2026doubleadiabaticequationsstaterelativistic} are similar but they use general symmetry principles in phase space to obtain these results.

In contrast, our model preserves full generality by deriving the evolution equations directly from the covariant Vlasov-Maxwell system without assuming homogeneity or non-relativistic flow. 
Our closure relationship relies on an anisotropic distribution function and we employ relativistic linear response theory to self-consistently incorporate heat fluxes driven by Landau damping.
This allows our framework to describe the dissipative kinetic effects that are not analyzed in~\citet{ley2026doubleadiabaticequationsrelativisticregime,wierzchucka2026doubleadiabaticequationsstaterelativistic}.
\section{Conclusions}\label{sec:conclusions}
Diffuse accretion flows are fundamentally weakly collisional or collisionless, rendering traditional models—such as the ideal MHD approximation or the collisional Braginskii regime—(formally) physically insufficient. In a weakly collisional plasma, analyzing the interplay among strong magnetic fields, anisotropic pressure transport, and non-local heat fluxes driven by particle kinetics is important for faithfully modeling the impact of microscopic scales on macroscopic flow dynamics.
In the non-relativistic limit, a fundamental model for the macroscopic dynamics of a strongly magnetized, weakly collisional plasma is KMHD~\citep{1983bpp..conf....1K}.
In this paper, we have, for the first time, extended the KMHD framework to general relativity. Our results provide a framework for modeling the impact of ion-electron kinetics in global simulations of accretion disks in curved spacetime, complementing the more fundamental but expensive PIC simulations~\citep{Galishnikova_2023}. 

We have also presented a general strategy to obtain a Landau fluid closure model for relativistic flows~\cite{1997PhPl....4.3974S}. In particular, we have presented analytic closure relationships for the heat flow parallel and perpendicular to the magnetic field for ultra-relativistic equation of state. 
The heat fluxes in the ultra-relativistic regime differ functionally from the non-relativistic heat flux due to the difference in the behavior of the Maxwell-Juttner distribution in the ultra-relativistic regime compared to the non-relativistic regime. 
A comparison between the linear response of the fluid equations and that of the kinetic equations shows that our closure relationship qualitatively captures the Landau damping predicted by the kinetic equations.

The theoretical framework presented in this paper enables many interesting directions for numerical methods to explore of the impact of pressure anisotropy in weakly collisional relativistic flows. 
However, devising numerical schemes to evolve the pressure anisotropy model proposed in this paper will be challenging.
For non-relativistic flows, the presence of the first and second adiabatic invariants helps one obtain a system of hyperbolic conservation laws for the Landau fluid system~\citep{Squire2023}.
This simplification is not possible in special- or general-relativistic flows.
A stable numerical scheme in the relativistic regime might require adopting the techniques used in the evolution of viscous neutron star mergers~\citep{Chabanov:2023blf}, given the similarities in the mathematics of these two problems.

It would be interesting to compare the results from our model to the Braginskii-like model used in~\citep{Chandra_2015,Foucart_2015,Foucart_2017} and the non-relativistic simulations of~\citet{Sharma_2006,2007ApJ...667..714S} to relativistic accretion flow around a Kerr black hole. This would provide a closer comparison to kinetic particle-in-cell simulations of~\citet{Galishnikova_2023} than the ideal MHD model. Our model also allows one to study the role of mirror and firehose instabilities on diffuse accretion flows. By identifying the threshold for these instabilities in the relativistic regime, our framework could offer a better understanding of how particle kinetics can dictate the rates of angular momentum transport and energy dissipation near the horizons of SMBHs. 
Another natural extension of this study would be to move toward higher-order moment closures~\citep{2020PhPl...27h2106N} or construct generalized parallel-kinetic and perpendicular moment models~\citep{Juno_2025} to accurately match the kinetic response function. Finally, one could investigate how the inclusion of pressure anisotropy modifies the growth rates of the MRI in rotating relativistic fluids, generalizing the non-relativistic study from~\citet{Quataert_2002,Sharma_2003}. 
\section*{Acknowledgements}
The authors thank Charles Gammie, James Juno, Matthew W. Kunz, Frans Pretorius, Eliot Quataert, and Jason M. TenBarge for illuminating discussions.
We are grateful to Charles Gammie and James Juno for reading the draft and providing detailed comments, which improved the presentation of this paper.

\section*{Declaration of Interests}
The authors report no conflict of interest.

\input{appendix}
\bibliographystyle{jpp}
\bibliography{ref}
\end{document}

%% file: appendix.tex
\appendix
\section{Derivation of the drift kinetic equation}\label{appendix:derivation-gyrokinetic}
\subsection{Deriving $\partial f_{0,s}/\partial \vartheta = 0$}
The first step in the analysis is to change coordinates from $(x,k^{\mu})$ to $(x,W_U, v_{\parallel}, v_{\perp}, \vartheta)$.
Using Eq.~\eqref{eq:kmu-decomposition}, we can show that
\begin{subequations}\label{eq:jacbian-V-k}
\begin{align}
    &\frac{\partial W_{U}}{\partial k^{\mu}} = - U_{\mu} \,,
    \quad 
    \frac{\partial v_{\parallel}}{\partial k^{\mu}} = \hat{b}_{\mu} \,,
    \,\\
    &\frac{\partial v_{\perp}}{\partial k^{\mu}} = \left( \cos(\vartheta) \hat{X}_{\mu} + \sin(\vartheta) \hat{Y}_{\mu} \right) \,, \quad
    \frac{\partial \vartheta}{\partial k^{\mu}} = - \frac{1}{v_{\perp}}\left( \sin \vartheta \hat{X}_{\mu} - \cos \vartheta \hat{Y}_{\mu} \right)\,.
\end{align}
\end{subequations}
Next, we use the decomposition of the electromagnetic tensor
\begin{align}\label{eq:Fmunu-decomposition}
    F^{\mu \nu} = U^{\mu} E^{\nu} - E^{\mu} U^{\nu} + 
    \frac{1}{2} U_{\delta} \epsilon^{\delta \mu \nu \gamma} b_{\gamma} \,.
\end{align}
To leading order $E_{\mu} = 0$ [Eq.~\eqref{eq:degeneracy-condition}]. Hence, using Eqs.~\eqref{eq:jacbian-V-k} and \eqref{eq:Fmunu-decomposition}, Eq.~\eqref{eq:vlasov-zeroth-order-GR} simplifies to
\begin{align}
    F^{\alpha}{}_{\mu} k^{\mu} \frac{\partial f_{s,0}}{\partial k^{\alpha}}
    \propto
    \frac{\partial f_{s,0}}{\partial \vartheta}
    =
    0\,.
\end{align}
\subsection{Drift kinetic equation}
We now derive the drift kinetic Vlasov equations. The velocity components of the Jacobian for the coordinate change were listed in Eq.~\eqref{eq:jacbian-V-k}. The coordinate space components are
\begin{subequations}\label{eq:jacobian-V-x}
\begin{align}
    &\frac{\partial W_{U}}{\partial x^{\beta}} = - k^{\mu} \frac{\partial U_{\mu}}{\partial x^{\beta}} \,, \quad
    \frac{\partial v_{\parallel}}{\partial x^{\beta}} = k^{\mu} \frac{\partial \hat{b}_{\mu}}{\partial x^{\beta}} \,,\\
    &\frac{\partial v_{\perp}}{\partial x^{\beta}} = \left( \cos(\vartheta) k^{\mu} \frac{\partial \hat{X}_{\mu}}{\partial x^{\beta}} + \sin(\vartheta) k^{\mu} \frac{\partial \hat{Y}_{\mu}}{\partial x^{\beta}} \right) \,, 
    \quad
    \frac{\partial \vartheta}{\partial x^{\beta}} = 
    \frac{k^{\mu}}{v_{\perp}} 
    \left[
    \cos \vartheta  \frac{\partial \hat{Y}_{\mu}}{\partial x^{\beta}}
    -
    \sin \vartheta \frac{\partial \hat{X}_{\mu}}{\partial x^{\beta}}
    \right]
    \,.
\end{align}
\end{subequations}
Using Eqs.~\eqref{eq:jacbian-V-k} and \eqref{eq:jacobian-V-x}, we have that
\begin{subequations}
\begin{align}
    &\frac{\partial f}{\partial x^{\beta}}
    =
    \frac{\partial f}{\partial x^{\beta}}
    +
    \frac{\partial f}{\partial W_{U}} \frac{\partial W_U}{\partial x^{\beta}}
    +
    \frac{\partial f}{\partial v_{\parallel}} \frac{\partial v_{\parallel}}{\partial x^{\beta}}
    + 
    \frac{\partial f}{\partial v_{\perp}}
    \frac{\partial v_{\perp}}{\partial x^{\beta}}
    +
    \frac{\partial f}{\partial \vartheta} 
    \frac{\partial \vartheta}{\partial x^{\beta}} \,, \nonumber\\
    &\overset{\mathrm{OMS}}{=}
    \frac{\partial f}{\partial x^{\beta}}
    +
    \frac{1}{W_U}\frac{\partial f}{\partial W_U}
    \left[ 
    W_U \frac{\partial W_U}{\partial  x^{\beta}}
    -
    v_{\parallel} \frac{\partial v_{\parallel}}{\partial x^{\beta}}
    -
    v_{\perp} \frac{\partial v_{\perp}}{\partial x^{\beta}}
    \right]
    \nonumber\\
    &
    +
    \frac{\partial f}{\partial v_{\parallel}} \left[k^{\mu} \frac{\partial \hat{b}_{\mu}}{\partial x^{\beta}} \right]
    + 
    \frac{\partial f}{\partial v_{\perp}}
    \left( \cos(\vartheta) k^{\mu} \frac{\partial \hat{X}_{\mu}}{\partial x^{\beta}} + \sin(\vartheta) k^{\mu} \frac{\partial \hat{Y}_{\mu}}{\partial x^{\beta}} \right)
    \nonumber\\
    &
    +
    \frac{k^{\mu}}{v_{\perp}}
    \frac{\partial f}{\partial \vartheta} 
    \left[
    \cos \vartheta  \frac{\partial \hat{Y}_{\mu}}{\partial x^{\beta}}
    -
    \sin \vartheta \frac{\partial \hat{X}_{\mu}}{\partial x^{\beta}}
    \right] \,,\\
    &\overset{\mathrm{OMS}}{=}
    \frac{\partial f}{\partial x^{\beta}}
    +
    \frac{\partial f}{\partial v_{\parallel}} \left[k^{\mu} \frac{\partial \hat{b}_{\mu}}{\partial x^{\beta}} \right]
    + 
    \frac{\partial f}{\partial v_{\perp}}
    \left( \cos(\vartheta) k^{\mu} \frac{\partial \hat{X}_{\mu}}{\partial x^{\beta}} + \sin(\vartheta) k^{\mu} \frac{\partial \hat{Y}_{\mu}}{\partial x^{\beta}} \right)
    \nonumber\\
    &
    +
    \frac{k^{\mu}}{v_{\perp}}
    \frac{\partial f}{\partial \vartheta} 
    \left[
    \cos \vartheta  \frac{\partial \hat{Y}_{\mu}}{\partial x^{\beta}}
    -
    \sin \vartheta \frac{\partial \hat{X}_{\mu}}{\partial x^{\beta}}
    \right] \,,\\
    &\frac{\partial f}{\partial k^{\beta}}
    =
    \frac{\partial f}{\partial W_{U}}  
    \frac{\partial W_U}{\partial k^{\beta}}
    +
    \frac{\partial f}{\partial v_{\parallel}} \frac{\partial v_{\parallel}}{\partial k^{\beta}}
    +
    \frac{\partial v_{\perp}}{\partial k^{\beta}}
    \frac{\partial f}{\partial v_{\perp}} 
    +
    \frac{\partial \vartheta}{\partial k^{\beta}}
    \frac{\partial f}{\partial \vartheta}  
    \,,
    \nonumber\\
    &\overset{\mathrm{OMS}}{=}
    \frac{\partial f}{\partial v_{\parallel}} \hat{b}_{\beta}
    +
    \bigg[\left( \cos(\vartheta) \hat{X}_{\mu} + \sin(\vartheta) \hat{Y}_{\mu} \right)\bigg]
    \frac{\partial f}{\partial v_{\perp}} 
    +
    \bigg[\frac{1}{v_{\perp}}\left( \cos \vartheta \hat{Y}_{\mu} - \sin \vartheta \hat{X}_{\mu}  \right)\bigg]
    \frac{\partial f}{\partial \vartheta} 
    \,,\\
    &F^{\alpha}{}_{\mu} k^{\mu}
    =
    U^{\alpha} \left( E_{\mu} v^{\mu} \right)
    +
    E^{\alpha} W_{U}
    +
    U_{\delta} \epsilon^{\delta \alpha \mu \gamma} b_{\gamma} v^{\perp}_{\mu} \,,
\end{align}
\end{subequations}
where $\mathrm{OMS}$ is a shorthand for on mass shell. To obtain the equations on shell we have used
\begin{align}
    \left.\frac{\partial f}{\partial v_{\parallel,\perp}}\right|_{W_{U}}
    =
    \frac{\partial f}{\partial v_{\parallel,\perp}}
    -
    \left.\frac{\partial f}{\partial W_U}\right|_{v_{\parallel,\perp}} \frac{v_{\parallel,\perp}}{W_U}
    \,.
\end{align}
With these identities the Vlasov equations for particle $s$ simplify to
\begin{align}
    &k^{\beta}
    \frac{\partial f}{\partial x^{\beta}}
    +
    k^{\beta}k^{\mu}\frac{\partial f}{\partial v_{\parallel}}  \frac{\partial \hat{b}_{\mu}}{\partial x^{\beta}}
    \nonumber\\
    &
    +
    k^{\beta}k^{\mu}
    \bigg[ 
    \frac{\partial f}{\partial v_{\perp}}
    \left( \cos(\vartheta) \frac{\partial \hat{X}^{\mu}}{\partial x^{\beta}} + \sin(\vartheta) \frac{\partial \hat{Y}^{\mu}}{\partial x^{\beta}} \right)
    +
    \frac{1}{v_{\perp}}
    \frac{\partial f}{\partial \vartheta} 
    \left[
    \cos \vartheta  \frac{\partial \hat{Y}^{\mu}}{\partial x^{\beta}}
    -
    \sin \vartheta \frac{\partial \hat{X}^{\mu}}{\partial x^{\beta}}
    \right]
    \bigg]
    \nonumber\\
    &+
    \left[
    \frac{q}{m} F^{\alpha}{}_{\mu} k^{\mu} 
    -
    \Gamma^{\alpha}{}_{\mu \nu} k^{\mu} k^{\nu}
    \right]
    \frac{\partial f}{\partial k^{\alpha}}
    =C[f]\,,
\end{align}
where we have dropped the species label $s$ for simplicity.
We can replace partial derivatives by covariant derivatives to obtain
\begin{align}
    &k^{\beta}
    \frac{\partial f}{\partial x^{\beta}}
    +
    k^{\beta}k^{\mu}\frac{\partial f}{\partial v_{\parallel}}  \nabla_{\beta} \hat{b}_{\mu}
    +
    \nonumber\\
    &
    +
    k^{\beta}k^{\mu}
    \bigg[ 
    \frac{\partial f}{\partial v_{\perp}}
    \left( \cos(\vartheta)\nabla_{\beta} \hat{X}_{\mu}  + \sin(\vartheta) \nabla_{\beta} \hat{Y}_{\mu} \right)
    +
    \frac{1}{v_{\perp}}
    \frac{\partial f}{\partial \vartheta} 
    \left[
    \cos \vartheta  \nabla_{\beta} \hat{Y}_{\mu}
    -
    \sin \vartheta \nabla_{\beta} \hat{X}_{\mu}
    \right]
    \bigg]
    \nonumber\\
    &+
    \left[
    \frac{q}{m} F^{\alpha}{}_{\mu} k^{\mu} 
    \right]
    \frac{\partial f}{\partial k^{\alpha}}
    =C[f]\,.
\end{align}
Observe that the Christoffel symbols no longer enter the equation.
To obtain the gyrokinetic equation, we replace $f\to f_0$, use Eq.~\eqref{eq:gyrotropic-plasma-condition} and take the average of the above equation over $\vartheta$ to get
\begin{subequations}\label{eq:gyrotropic-eqn-id-1}
\begin{align}
    &\left<\frac{q}{m} F^{\alpha}{}_{\mu} k^{\mu} \frac{\partial f}{\partial k^{\alpha}} \right>_{\vartheta}
    =
    \frac{q}{m} E_{\parallel} W_{U} \frac{\partial f_0}{\partial v_{\parallel}} \,\\
    &\left< k^{\beta} \frac{\partial f}{\partial x^{\beta}} \right>_{\vartheta}
    =
    W_{U} U^{\beta} \frac{\partial f_0}{\partial x^{\beta}}
    +
    v_{\parallel} \hat{b}^{\beta} \frac{\partial f_0}{\partial x^{\beta}}
    \,, 
    \\
    &\left<
    k^{\beta}k^{\mu}
    \left( \cos(\vartheta)\nabla_{\beta} \hat{X}_{\mu}  + \sin(\vartheta) \nabla_{\beta} \hat{Y}_{\mu}  \right)
    \right>_{\vartheta}
    =
    \tfrac{1}{2} (a_{U}^{\alpha } \hat{b}_{\alpha } -  \nabla_{\alpha} \hat{b}^{\alpha}) v_{||}{} v_{\perp}{} 
    \nonumber\\
    &+ \tfrac{1}{2} v_{\perp} W_{U} (- \nabla_{\alpha} U^{\alpha} + \hat{b}^{\alpha } \hat{b}^{\beta } \nabla_{\beta }U_{\alpha }) \,,\\
    &\left<k^{\beta}k^{\mu}\nabla_{\beta} \hat{b}_{\mu} \right>_{\vartheta}
    =
    \tfrac{1}{2} (- a_{U}^{\alpha } \hat{b}_{\alpha } + \nabla_{\alpha} \hat{b}^{\alpha}) v_{\perp}^2 -  a_{U}^{\alpha } \hat{b}_{\alpha } W_{U}{}^2 -
    v_{||} \hat{b}^{\alpha } \hat{b}^{\beta } W_{U} \nabla_{\beta }U_{\alpha }
    \,,
\end{align}
\end{subequations}
where $a_{U}^{\alpha} = D U^{\alpha}$ is the acceleration.
We can simplify this further by noting that for magnetically dominated electromagnetic field tensor, the Gauss-Faraday law yields,
\begin{subequations}\label{eq:gyrotropic-eqn-id-2}
\begin{align}
    &b_{\beta} \nabla_{\alpha} \left(\star F^{\alpha \beta} \right) = 0 \implies 
     \hat{b}^{\alpha } \hat{b}^{\beta } \nabla_{\beta }U_{\alpha }
     - \nabla_{\alpha} U^{\alpha}
     =
     \frac{D b}{b} \,,\\
     &U_{\beta} \nabla_{\alpha} \left(\star F^{\alpha \beta} \right) = 0 \implies
     \hat{b}^{\beta} a_{\beta}^U - \nabla_{\alpha} \hat{b}^{\alpha} = \frac{\hat{b}^{\beta}}{b} \nabla_{\beta} b \,.
\end{align}  
\end{subequations}
Therefore, we have
\begin{subequations}\label{eq:gyrotropic-eqn-id-3}
\begin{align}
    &\left<
    k^{\beta}k^{\mu}
    \left( \cos(\vartheta)\nabla_{\beta} \hat{X}_{\mu}  + \sin(\vartheta) \nabla_{\beta} \hat{Y}_{\mu} \right)
    \right>_{\vartheta}
    =
    \frac{W_{U} v_{\perp}}{2 b} D_{\parallel} b
    \\
    &\left<k^{\beta}k^{\mu}\nabla_{\beta} \hat{B}^U_{\mu} \right>_{\vartheta}
    =
    - W_{U}^2 \hat{b}^{\beta} D_{\parallel} U_{\beta}
    -  \frac{\hat{b}^{\alpha } v_{\perp}{}^2 \nabla_{\alpha }b}{2 b}
    \,.
\end{align}
\end{subequations}
Combining Eqs.~\eqref{eq:gyrotropic-eqn-id-1}-\eqref{eq:gyrotropic-eqn-id-3} we obtain Eq.~\eqref{eq:gyroaveraged-kinetic-equation}.
\subsection{Volume element}
The volume element over the mass shell is given by~\citep{Sarbach2013}
\begin{align}
    d V \equiv \sqrt{-g} \, d^4 k
    =
    2\sqrt{-g}
    \,
    d^4 k
    \,
    \delta \left( k_{\mu} k^{\mu} + 1 \right) \theta( k_{\mu} n^{\mu})
    \,.
\end{align}
We change coordinates from $k^{\mu} \to \Vec{V} \equiv \left(W_U, v_{\parallel}, v_{\perp}, \vartheta\right)$,
\begin{align}
    d V = \sqrt{-g} \, d^4 k
    =
    \sqrt{-g}
    \,
    \left|
    \mathrm{det}
    \left[\frac{\partial k^{\mu}}{\partial \Vec{V} } \right]
    \right|
    d W_U d v_{\perp} d v_{\parallel} d \vartheta
    \,.
\end{align}
The Jacobian can be calculated using Eq.~\eqref{eq:jacbian-V-k}.
Define a tetrad
\begin{align}
    e^{a}{}_{\mu}
    \equiv
    \left(U_{\mu}, \hat{b}_{\mu}, \hat{X}_{\mu}, \hat{Y}_{\mu} \right)
    \,.
\end{align}
We know that
\begin{align}
    g_{\mu \nu} = \eta_{a b} e^{a}{}_{\mu} e^{b}{}_{\nu} \implies
    \mathrm{det} \left[e^{a}{}_{\mu} \right]
    =
    \sqrt{-g}
    \,.
\end{align}
From Eq.~\eqref{eq:jacbian-V-k}, we see that
\begin{align}
    \frac{\partial \Vec{V}^{a}}{\partial k^{\mu}}
    =
    \boldsymbol{E}^{a}{}_{b} e^{b}{}_{\mu}\,,
    \quad
    \boldsymbol{E}^{a}{}_{b}
    =
    \begin{pmatrix}
        -1 & 0 & 0 & 0 \\
        0 & 1 & 0 & 0 \\
        0 & 0 & \cos \vartheta & \sin \vartheta \\
        0 & 0 & - v_{\perp}^{-1}\sin \vartheta & v_{\perp}^{-1}\cos \vartheta
    \end{pmatrix}
    \,.
\end{align}
Therefore,
\begin{align}
    \left|\mathrm{det}
    \left[ \frac{\partial \Vec{V}^{a}}{\partial k^{\mu}}\right]
    \right|
    =
    \sqrt{-g}
    \left|\mathrm{det} \left[\boldsymbol{E}^{a}{}_{b} \right]
    \right|
    =
    \frac{1}{v_{\perp}} \sqrt{-g}
    \,.
\end{align}
Combining the above equations, we obtain that
\begin{align}
    d V = \sqrt{-g} \, d^4 k
    =
    v_{\perp}
    d W_U d v_{\perp} d v_{\parallel} d \vartheta
    \overset{\mathrm{OMS}}{=} \frac{v_{\perp} d v_{\perp} d v_{\parallel} d \vartheta}{\sqrt{1+ v_{\perp}^2 + v_{\parallel}^2}}
    \,.
\end{align}
\section{Evaluating the moments of the anisotropic distribution function}\label{appendix:moments-of-distribution-function}
For simplicity, in this section, we drop the species label $s$. 
The moments $\mathcal{I}^{(P,Q,R)}$ [Eq.~\eqref{eq:IPQ-def}] are given by
\begin{align}\label{eq:IPQR-appendix-1}
    \mathcal{I}^{(P,Q,R)}
    =
    2\pi \alpha
    \int_{v_{\perp}=0}^{\infty}
    \int_{v_{\parallel}=-\infty}^{\infty}
    d v_{\perp} d v_{\parallel}
    v_{\perp}^{P+1} v_{\parallel}^{Q} \left(
    \sqrt{1+v_{\parallel}^2 + v_{\perp}^2}\right)^{R-1}
    e^{-z\sqrt{1 + v_{\parallel}^2 \delta_1 + v_{\perp}^2 \delta_2}}
    \,.
\end{align}
In this paper, we are interested in the ultra-relativistic limit.
This limit becomes transparent if we use the substitution 
\begin{align}
    v_{\perp} = \frac{w}{z} \cos(\theta) \,, \quad 
    v_{\parallel} = \frac{w}{z} \sin(\theta) \,,
\end{align}
to reduce the integral to
\begin{align}
    \mathcal{I}^{(P,Q,R)}
    &=
    \frac{2\pi \alpha}{z^{2+P+Q+R}}
    \int_{w=0}^{\infty}
    \int_{\theta = 0}^{\pi}
    dw d\theta
    \,
    w^{P+Q+2} \sin ^{P+1}(\theta ) \cos ^Q(\theta ) \left(w^2+z^2\right)^{\frac{R-1}{2}} \nonumber\\
    &\times
    e^{-\sqrt{w^2 \left(\delta_1 \cos ^2(\theta )+\delta_2 \sin ^2(\theta )\right)+z^2}}
    \nonumber\\
    &
    \equiv
    \frac{2\pi \alpha}{z^{2+P+Q+R}}
    \tilde{\mathcal{I}}^{(P,Q,R)}
    \,.
\end{align}
Similarly we can reduce $\mathcal{I}^{(-1)}_{(P,Q,R)}$ [Eq.~\eqref{eq:Iminus-moment}] to
\begin{align}
    &\mathcal{I}^{(-1)}_{(P,Q,R)}
    =
    \frac{2 \pi \alpha}{z^{1+P+Q+R}}
    \nonumber\\
    &\int_{w=0}^{\infty}
    \int_{\theta = 0}^{\pi}
    dw d\theta \,
    \frac{w^{P+Q+2} \sin ^{P+1}(\theta ) \cos^Q(\theta ) \left(w^2+z^2\right)^{\frac{R-1}{2}} e^{-\sqrt{w^2 \left(\delta_1 \cos ^2(\theta )+\delta_2 \sin ^2(\theta )\right)+z^2}}}{\sqrt{w^2 \left(\delta_1 \cos ^2(\theta )+\delta_2 \sin ^2(\theta )\right)+z^2}}
    \nonumber\\
    &
    \equiv
    \frac{2 \pi \alpha}{z^{1+P+Q+R}}
    \tilde{\mathcal{I}}^{(-1)}_{(P,Q,R)}
    \,.
\end{align}
For a given value of $(z,\delta_1,\delta_2)$, the integrals $\tilde{\mathcal{I}}^{(P,Q,R)}$ and $\tilde{\mathcal{I}}^{(-1)}_{(P,Q,R)}$ can be evaluated numerically. When $P+Q+R<0$, it is useful to switch the integration variable to
\begin{align}
    w \to \sqrt{w^2-z^2}
\end{align}
to ensure better convergence.
\subsection{Analytical expression for the moments in the high-temperature limit}\label{appendix:analytical-moments}
Define
\begin{align}\label{eq:mutilde-Delta-tilde}
    \tilde{\mu} \equiv \frac{\delta_1}{\delta_2}-1\,,
    \tilde{\Delta} \equiv \frac{\delta_2-1}{\delta_1-1} \,.
\end{align}
To yield physical values for thermodynamic variables, we require that $\infty>\tilde{\mu}>-1$. In the high-temperature limit, one can evaluate the moments analytically using Eq.~\eqref{eq:IPQR-appendix-1}.
Some useful moments are listed here
\begin{subequations}\label{eq:moments-distribution-function-analytical-appendix}
\begin{align}
    n &= \frac{8 \pi  \alpha }{\sqrt{\tilde{\mu }+1} \delta _2{}^{3/2} z^3}\,,
    \quad
    \mathcal{E} = 2 p_{\perp} + p_{\parallel} 
    \\
    p_{\parallel} &= 
    \frac{12 \pi  \alpha  m \tan ^{-1}\left(\sqrt{\tilde{\mu }}\right)}{\tilde{\mu }^{3/2} \delta _2{}^2 z^4}-\frac{12 \pi  \alpha  m_s}{\tilde{\mu } (\tilde{\mu }+1) \delta _2{}^2 z^4}
    \,, \\
    p_{\perp} &= 
    \frac{6 \pi  \alpha  (\tilde{\mu }-1) m \tan ^{-1}\left(\sqrt{\tilde{\mu }}\right)}{\tilde{\mu }^{3/2} \delta _2{}^2 z^4}+\frac{6 \pi  \alpha  m}{\tilde{\mu } \delta _2{}^2 z^4}
    \,,\\
    r_{\parallel,\parallel}^{(-2)}
    &=
    \frac{12 \pi  \alpha  (2 \tilde{\mu }+3) m_s}{\tilde{\mu }^2 (\tilde{\mu }+1) \delta _2{}^2 z^4}-\frac{36 \pi  \alpha  m_s \tan ^{-1}\left(\sqrt{\tilde{\mu }}\right)}{\tilde{\mu }^{5/2} \delta _2{}^2 z^4}\,\\
    r_{\parallel,\perp}^{(-2)}
    &=
    \frac{6 \pi  \alpha  (\tilde{\mu }+3) m_s \tan ^{-1}\left(\sqrt{\tilde{\mu }}\right)}{\tilde{\mu }^{5/2} \delta _2{}^2 z^4}-\frac{18 \pi  \alpha  m_s}{\tilde{\mu }^2 \delta _2{}^2 z^4}
    \,,\\
    r_{\perp,\perp}^{(-2)}
    &=
    \frac{3 \pi  \alpha  (\tilde{\mu }-3) (\tilde{\mu }+1) m_s \tan ^{-1}\left(\sqrt{\tilde{\mu }}\right)}{\tilde{\mu }^{5/2} \delta _2{}^2 z^4}+\frac{3 \pi  \alpha  (\tilde{\mu }+3) m_s}{\tilde{\mu }^2 \delta _2{}^2 z^4}
    \,.
\end{align}
\end{subequations}
Finally, we note that the anisotropic distribution function [Eq.~\eqref{eq:anisotropic-distribution-function-relativity}] contains four free parameters $(\delta_{1,2}, \alpha, z)$.
We can fix of one of these parameters by demanding that the anisotropic distribution function satisfies the ideal gas law
\begin{align}\label{eq:fix-ideal-gas-law}
    \frac{2}{3} p_{\perp} + \frac{1}{3} p_{\parallel} 
    =
    \frac{m n}{z}
\end{align}
which leads to the following equation for $\tilde{\Delta}$ 
\begin{align}
    \tilde{\Delta}
    =
    \left(1+\tilde{\mu} \right)^{-1}
    \left[
    \frac{1}{\frac{1}{\tilde{\mu }}-\frac{(\tilde{\mu }+1) \tan ^{-1}\left(\sqrt{\tilde{\mu }}\right)}{\tilde{\mu }^{3/2}}}+\frac{1}{\frac{(\tilde{\mu }+1) \tan ^{-1}\left(\sqrt{\tilde{\mu }}\right)}{\tilde{\mu }^{3/2}}+\frac{3}{\tilde{\mu }}}+1
    \right]
    \,.
\end{align}
Observe that the relation used in Eq.~\eqref{eq:fix-ideal-gas-law} is not a non-relativistic equation of state. Since $z\sim m/T$, the right-hand side is $\sim n T$ . 
Hence, Eq.~\eqref{eq:fix-ideal-gas-law} fixes the effective temperature parameter of the anisotropic distribution function.
\section{Relativistic plasma response function}\label{appendix:plasma-response-function}
In this appendix, we discuss how to evaluate $\mathcal{H}_s^{(P,Q,R)}$. Let us start by proving Eq.~\eqref{eq:H-zeta-expansion}. For $n>0$, we note the following elementary factorization identity
\begin{align}\label{eq:elementary-identity}
    \lambda^n - \zeta^n = (\lambda-\zeta)(\lambda^{n-1} + \lambda^{n-2} \zeta + \ldots + \lambda \zeta^{n-2} + \zeta^{n-1})
    \,.
\end{align}
Ignore the species label $s$ for simplicity and rewrite Eq.~\eqref{eq:Hfunc-definition} as
\begin{align}
    \mathcal{H}^{(P,Q,R)}(\zeta)
    &=
    -z \delta_{1}
    \left<\frac{v_{\perp}^{P} v_{\parallel}^{Q+1} W_U^{R} }{ \left(v_{\parallel} W_{U}^{-1} - \zeta\right) \Tilde{W}} \right>
    =
    -z \delta_{1}
    \left<\frac{v_{\perp}^{P} \left(\lambda\right)^{Q+1} W_U^{R+Q+1} }{ \left(\lambda - \zeta\right) \Tilde{W}} \right> \,,\nonumber\\
\end{align}
where $\lambda \equiv v_{\parallel} W_U^{-1}$. Using Eq.~\eqref{eq:elementary-identity} and simplifying we obtain Eq.~\eqref{eq:H-zeta-expansion}.

\subsection{Evaluating $\mathcal{Z}^{(P,R)}$}
In this section, we assume that $P+R+1>0$.
From Eq.~\eqref{eq:Z-definition}, we have
\begin{align}
    \mathcal{Z}^{(P,R)} &= 
    2\pi \alpha
    \int_{v_{\perp}=0}^{\infty} v_{\perp} d v_{\perp}
    \int_{v_{\parallel}=-\infty}^{\infty} \frac{d v_{\parallel}}{W_U}
    \frac{v_{\perp}^P W_U^R}{\left(v_{\parallel} W_{U}^{-1} - \zeta\right) \Tilde{W} }
    e^{-z \tilde{W}}
    \,,\\
    &\overset{v = v_{\parallel} W_U^{-1}}{\Large{=}}
    2 \pi \alpha \int_{v_{\perp}=0}^{\infty} v_{\perp}^{P} 
    \left[1 + v_{\perp}^2 \right]^{R/2}
    v_{\perp} d v_{\perp}
    \int_{v=-1}^{1} \frac{dv}{\left(1-v^2\right)^{R/2+1}}
    \frac{e^{-z \tilde{W}}}{\tilde{W}(v - \zeta)}
    \,.
\end{align}
When $\zeta$ is real and we can use Landau's prescription to obtain
\begin{align}\label{eq:ZPR-v1}
    \mathcal{Z}^{(P,R)}
    &=
    2 \pi \alpha
    \bigg[
    \int_{v_{\perp}=0}^{\infty} v_{\perp}^{P} 
    \left[1 + v_{\perp}^2 \right]^{R/2}
    v_{\perp} d v_{\perp}
    \mathrm{P.V.}
    \int_{v=-1}^{1} \frac{dv}{\left(1-v^2\right)^{R/2+1}}
    \frac{e^{-z \tilde{W}}}{\tilde{W}(v - \zeta)}
    \bigg]
    \nonumber\\
    &+
    2 i \pi^2 \alpha
    \bigg[
    \int_{v_{\perp}=0}^{\infty} d v_{\perp} v_{\perp}^{P+1} 
    \left[1 + v_{\perp}^2 \right]^{R/2}
    \frac{e^{-z \tilde{W}}}{\tilde{W}\left(1-\zeta^2\right)^{R/2+1}}
    \bigg]
    \,.
\end{align}
\subsubsection{Evaluation in the low-frequency limit}
In the low-frequency limit only the second term in Eq.~\eqref{eq:ZPR-v1} contributes. We can evaluate this by changing variables to
\begin{align}
    w = z v_{\perp}\,,
\end{align}
to obtain
\begin{align}
    \mathcal{Z}^{(P,R)}
    &=
    \frac{2 i \pi^2 \alpha }{z^{P+R+1}}
    \int_{w=0}^{\infty}
    dw\,
    w^{P+1} \left(w^2+z^2\right)^{R/2} \frac{e^{-\sqrt{\delta_2 w^2+z^2}}}{\sqrt{\delta_2 w^2+z^2}}
    +
    \mathcal{O}(\zeta)
    \nonumber\\
    &\equiv
    \frac{2 i \pi^2 \alpha}{z^{P+R+1}}
    \tilde{\mathcal{Z}}^{(P,R)}_{I}(\zeta=0)
    \,.
\end{align}
In the high-temperature limit, $\tilde{\mathcal{Z}}^{(P,R)}_{I}(\zeta=0)$ evaluates to $\delta_2^{-P/2-R/2-1} \Gamma(P+R+1)$.
\subsubsection{Evaluation for general a real value of $-1<\zeta<1$}
The second integral in Eq.~\eqref{eq:ZPR-v1} is purely imaginary and we can evaluate this using the transformations we have discussed so far
\begin{align}
    \mathrm{Im}\left[\mathcal{Z}^{(P,R)} \right]
    &=
    \frac{2 i \pi^2 \alpha}{z^{1+P+R} (1-\zeta^2)^{1+R/2}}
    \bigg[
    \int_{w=0}^{\infty}
    dw \,
    w^{P+1} \left(w^2+z^2\right)^{R/2} \nonumber\\
    &\times
    \frac{e^{-\sqrt{\delta_2 w^2-\frac{\delta_1 \zeta ^2 \left(w^2+z^2\right)}{\zeta ^2-1}+z^2}}}{\sqrt{\delta_2 w^2-\frac{\delta_1 \zeta ^2 \left(w^2+z^2\right)}{\zeta ^2-1}+z^2}}
    \bigg]
    \nonumber\\
    &\equiv
    \frac{2 i \pi^2 \alpha}{z^{1+P+R}}
    \tilde{\mathcal{Z}}_{I}^{(P,R)}(\zeta)
    \,.
\end{align}
In the high-temperature limit
\begin{align}
    \tilde{\mathcal{Z}}_{I}^{(P,R)}(\zeta)
    &=
    \frac{\Gamma(P+R+1)}{(1-\zeta^2)^{1+R/2}} \left[\delta_2 + \frac{\delta_1 \zeta^2}{1-\zeta^2} \right]^{-1-P/2-R/2}
    \,.
\end{align}
To obtain the real part of Eq.~\eqref{eq:ZPR-v1}, we analyze the principal value contribution
\begin{align}
    &\mathrm{P.V}
    \int_{v=-1}^{1} \frac{dv}{\left(1-v^2\right)^{R/2+1}}
    \frac{e^{-z \tilde{W}}}{\tilde{W}(v - \zeta)}
    =
    \frac{e^{-z \tilde{W}(v=\zeta)}}{\tilde{W}(v=\zeta) \left(1-\zeta^2\right)^{R/2+1}} \log\left[ \frac{|1-\zeta|}{|1+\zeta|} \right]
    \nonumber\\
    &+
    \int_{v=-1}^{1} \frac{dv}{v - \zeta}
    \left[
    \frac{e^{-z \tilde{W}}}{\tilde{W}\left(1-v^2\right)^{R/2+1}}
    -
    \frac{e^{-z \tilde{W}(v=\zeta)}}{\tilde{W}(v=\zeta)\left(1-\zeta^2\right)^{R/2+1}}
    \right]
    \,,
\end{align}
Substituting this in Eq.~\eqref{eq:ZPR-v1} and simplifying we obtain
\begin{align}\label{eq:ZPR-Re-1}
    &\mathrm{Re}\left[\mathcal{Z}^{(P,R)} \right]
    =
    \frac{2 \pi \alpha}{z^{1+P+R} } \log\left[ \frac{|1-\zeta|}{|1+\zeta|} \right]
    \tilde{\mathcal{Z}}_{I}^{(P,R)}
    \nonumber\\
    &+
    \frac{2\pi \alpha}{z^{P+R+1}}
    \int_{w=0}^{\infty} dw \int_{v=-1}^{1} dv
    \bigg[ 
    \frac{w^{P+1} \left(1-\zeta ^2\right)^{-\frac{R}{2}-1} \left(w^2+z^2\right)^{R/2} e^{-\sqrt{\delta_2 w^2-\frac{\delta_1 \zeta ^2 \left(w^2+z^2\right)}{\zeta ^2-1}+z^2}}}{(\zeta -v) (z \tilde{W})}
    \nonumber\\
    &-\frac{w^{P+1} \left(1-v^2\right)^{-\frac{R}{2}-1} \left(w^2+z^2\right)^{R/2} \exp \left(-\sqrt{-\frac{\delta_1 v^2 \left(w^2+z^2\right)}{v^2-1}+\delta_2 w^2+z^2}\right)}{(\zeta -v)(z \tilde{W})}
    \bigg]
    \,.
\end{align}
Define
\begin{align}
    y(\zeta) \equiv
    (1-\zeta^2)^{P/2} \left(\delta_2(1-v^2) + \delta_1 v^2\right)^{-1-P/2-R/2}
    \,.
\end{align}
In the high-temperature limit, the second integral in Eq.~\eqref{eq:ZPR-Re-1} reduces to
\begin{align}
    &\mathrm{Re}\left[\mathcal{Z}^{(P,R)} \right]
    =
    \frac{2 \pi \alpha}{z^{1+P+R} } 
    \log\left[ \frac{|1-\zeta|}{|1+\zeta|} \right]
    \tilde{\mathcal{Z}}_{I}^{(P,R)}
    +
    \frac{2\pi \alpha}{z^{P+R+1}}
    \int_{v=-1}^{1}
    dv
    \frac{y(v) - y(\zeta)}{v-\zeta}
    \,.
    \nonumber\\
    &=
    \frac{2 \pi \alpha}{z^{1+P+R} } 
    \log\left[ \frac{|1-\zeta|}{|1+\zeta|} \right]
    \tilde{\mathcal{Z}}_{I}^{(P,R)}
    \nonumber\\
    &+
    \frac{2\pi \alpha}{z^{P+R+1}}
    \lim_{\epsilon \to 0^{+}}
    \bigg[
    \int_{v=-1}^{-\epsilon}
    dv
    \frac{y(v) - y(\zeta)}{v-\zeta}
    +
    2 \epsilon\, y'(\zeta)
    +
    \int_{\epsilon}^{1}
    dv
    \frac{y(v) - y(\zeta)}{v-\zeta}
    \bigg]
    \,.
\end{align}
\section{Linearized fluid response}\label{appendix:linearized-fluid-and-MHD}
\begin{figure*}
    \centering
    \includegraphics[width=1\linewidth]{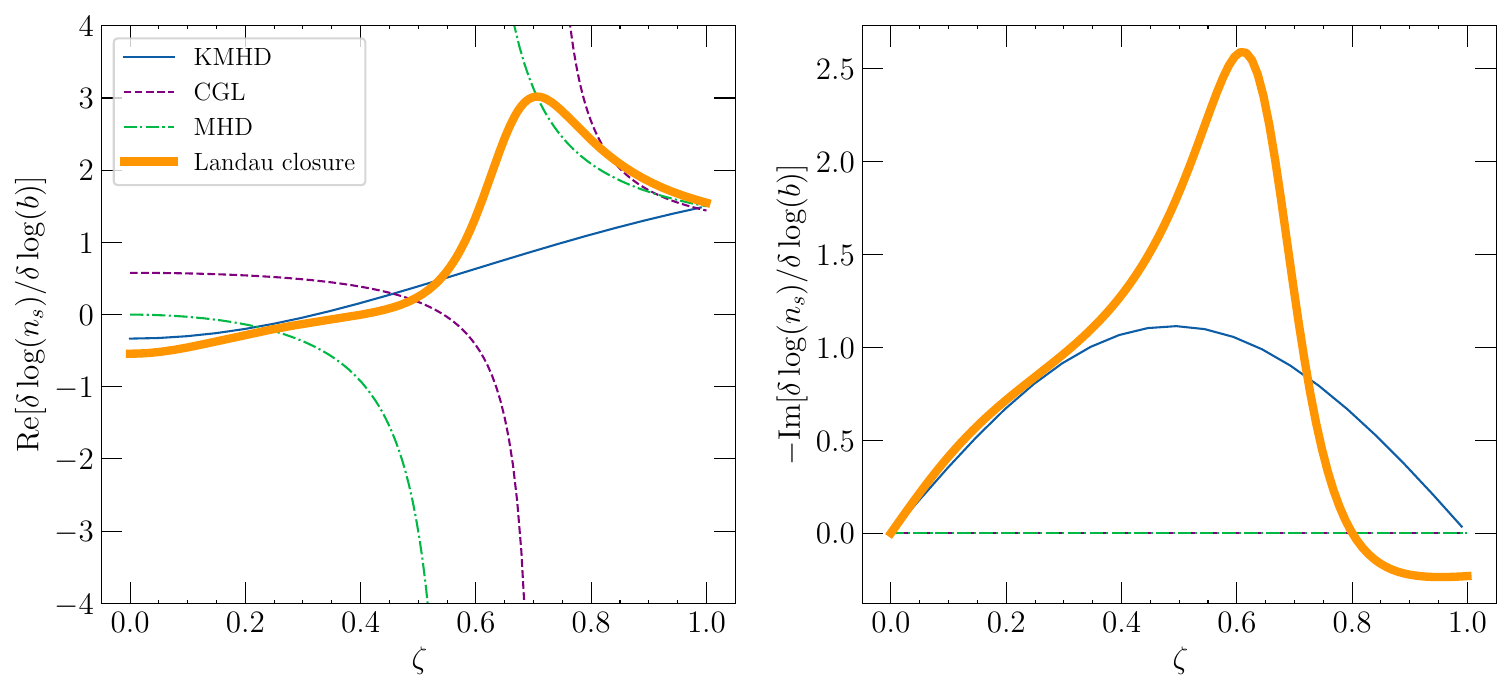}
    \includegraphics[width=1\linewidth]{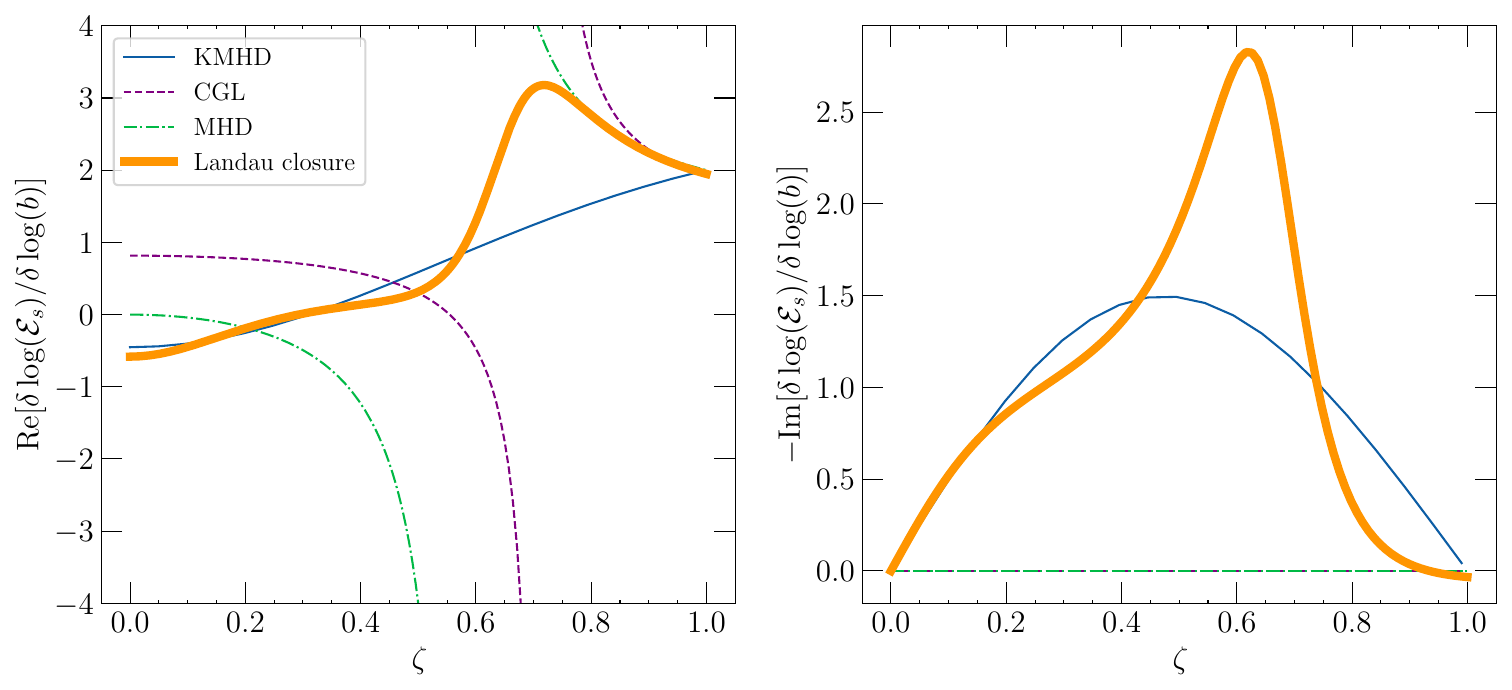}
    \caption{Comparison between the linear number density (top panel) and energy response (bottom panel) between the KMHD (solid blue line), CGL (dashed purple line), MHD (dash dotted green line) and the Landau fluid closure (bold orange line) as a function of the driving speed $\zeta = \omega/|k_z|$ for an initially anisotropic distribution function. 
    The response is qualitatively similar to the isotropic state, see Fig.~\ref{fig:response_comparison}.
    }
    \label{fig:response_comparison_aniso}
\end{figure*}
Here, we list the expressions for the fluid response functions plotted in Fig.~\ref{fig:response_comparison}.
The density response functions are
\begin{subequations}
\begin{align}
    &\left.\frac{\delta \log(n_s)}{\delta \log(b)}\right|_{\mathrm{MHD}}
    =
    \frac{3 \zeta ^2}{3 \zeta ^2-1}
    \,,\\
    &\left.\frac{\delta \log(n_s)}{\delta \log(b)}\right|_{\mathrm{CGL}}
    =
    \frac{2- 5 \zeta ^2}{\left(3-5 \zeta ^2\right)} \,,\\
    &\left.\frac{\delta \log(n_s)}{\delta \log(b)}\right|_{\mathrm{LF}}
    =
    \frac{\zeta  \left(3 \pi ^2 \left(5 \zeta ^2-2\right) \zeta +12 i \pi  \left(5 \zeta ^2-1\right)-64 \zeta \right)}{3 \pi ^2 \left(5 \zeta ^2-3\right) \zeta ^2-48 \zeta ^2+2 i \pi  \left(27 \zeta ^2-13\right) \zeta +16}
    \,.
\end{align}
\end{subequations}
The energy density response functions are
\begin{subequations}
\begin{align}
    &\left.\frac{\delta \log(\mathcal{E}_s)}{\delta \log(b)}\right|_{\mathrm{MHD}}
    =
    \frac{4 \zeta ^2}{3 \zeta ^2-1}
    \,,\\
    &\left.\frac{\delta \log(\mathcal{E}_s)}{\delta \log(b)}\right|_{\mathrm{CGL}}
    =
    \frac{8- 20 \zeta ^2}{9- 15 \zeta ^2} \,,\\
    &\left.\frac{\delta \log(\mathcal{E}_s)}{\delta \log(b)}\right|_{\mathrm{LF}}
    =
    \frac{4 \zeta  \left(\pi  \left(5 \zeta ^2-2\right)+8 i \zeta \right)}{3 \pi  \zeta  \left(5 \zeta ^2-3\right)+8 i \left(3 \zeta ^2-1\right)}
    \,.
\end{align}
\end{subequations}
\section{Comparison between kinetic and fluid responses for anisotropic initial state}\label{appendix:initially-anisotropic-distribution}
Here, we provide a comparison between the fluid and kinetic response functions for an initial anisotropic state. 
We let $\delta_{1} = 1.2$ and $\delta_2 = 0.9$ for the ions and electrons and this results in an initially anisotropy of 
\begin{align}
    \frac{\Delta p}{p} \approx 22 \% \,
\end{align}
for both the ions and electrons. The analytical expressions for the response functions in this case are complicated and we do not provide them here. 

The response functions are compared in Fig.~\ref{fig:response_comparison_aniso}. As in the isotropic case, the Landau fluid closure closely matches the kinetic response in the low-frequency limit and qualitatively reproduces the imaginary part of the response function. Unlike the isotropic state, the low-frequency response agreement also deteriorates slightly for the real part of the response functions. This is a consequence of the matching strategy which is designed to capture the imaginary part of the response due to Landau damping and phase mixing.
\section{Relativistic-fluid terminology}
\label{appendix:hydro-frames}
In this appendix, we briefly review some standard terminology from relativistic fluid dynamics that is used throughout this paper.
\subsection{Hydrodynamic decomposition and frame choice}
Most modern treatment of relativistic fluid dynamics~\cite{Israel:1979wp,Denicol_2012} begin by introducing an irreducible decomposition of the the particle current and stress energy tensor along a timelike fluid four-velocity $U^\mu$ [Eq.~\eqref{eq:number-density-and-stress-tensor-s-def}]
\begin{align*}
    N_s^\mu
    &=
    n_s U^\mu
    +
    V_s^\mu
    \,,\\
    T_s^{\mu\nu}
    &=
    \mathcal{E}_s U^\mu U^\nu
    +
    \mathcal{P}_s \Delta^{\mu\nu}
    +
    2\mathcal{Q}_s^{(\mu}U^{\nu)}
    +
    \pi_s^{\mu\nu}
    \,,
\end{align*}
where
\begin{align*}
    \Delta^{\mu\nu}
    =
    g^{\mu\nu}
    +
    U^\mu U^\nu
\end{align*}
projects orthogonal to $U^\mu$.
Here, $n_s$ is the particle number density, $V_s^\mu$ is the particle diffusion current, $\mathcal{E}_s$ is the energy density, $\mathcal{P}_s$ is the isotropic pressure, $\mathcal{Q}_s^\mu$ is the heat-flux vector, and $\pi_s^{\mu\nu}$ is the shear-stress tensor.
The terminology ``shear stress'' refers to the symmetric, trace-free part of the spatial stress tensor and does not neccesarily imply that the stress has a collisional or viscous origin.
For a gyrotropic plasma, $\pi_s^{\mu\nu}$ contains the pressure anisotropy $p_{\parallel,s}-p_{\perp,s}$, see Eq.~\eqref{eq:pimunu-gyrotropic}.
Similarly, particle diffusion current refers to the component of the particle current which is not parallel to $U^{\mu}$. This component need not be of viscous origin.

Up to this point, the time-like four velocity $U^{\mu}$ has been arbitrary. In fact, away from local thermodynamic equilibrium, the decomposition above does not uniquely define the fluid four-velocity $U^\mu$.
Fundamentally, this arises because the flow of energy and rest-mass density out-of-equilibrium need not be parallel to each other~\citep{Landau:1987}.
A ``hydrodynamic-frame choice'' specifies how $U^\mu$ is defined from the particle current or stress-energy tensor.
It specifies a form for the stress energy tensor and the particle currents; it does not make $U^\mu$ a fixed background vector.
Any choice is purely performed to make the subsequent calculation and interpretation easier.
For a survey of different hydrodynamic frame choices, see~\citet{Rocha-Denicol-Jorge-RTA}.

We now briefly describe the two most common hydrodynamic frame choices. 
The Eckhart frame used in this paper defines $U^{\mu}$ by demanding that the mass-weighted diffusion current vanishes,
\begin{align*}
    \sum_s m_s V_s^\mu
    =
    0
    \,.
\end{align*}
This choice is the relativistic analogue of the usual barycentric, or center-of-mass, velocity used in non-relativistic multi-fluid hydrodynamics. 
Indeed, in the non-relativistic limit, the spatial components of the condition above reduce to
\begin{align*}
    \sum_s \rho_s
    \left(
    \mathbf{v}_s-\mathbf{v}
    \right)
    =
    0
    \,,
\end{align*}
where $\rho_s=m_s n_s$ and $\mathbf{v}$ is the bulk fluid velocity. 
Equivalently,
\begin{align*}
    \rho\,\mathbf{v}
    =
    \sum_s \rho_s \mathbf{v}_s
    \,,
    \qquad
    \rho
    =
    \sum_s \rho_s
    \,,
\end{align*}
so that the bulk velocity is the mass-weighted average of the species velocities.

By contrast, in the Landau frame, the fluid velocity is chosen so that the total heat-flux vector vanishes,
\begin{align}
    \mathcal{Q}^\mu
    =
    \sum_{s} \mathcal{Q}_{s}^{\mu}
    =
    0
    \,.
\end{align}
This frame is commonly used in relativistic heavy-ion collisions, where at high collision energies the net conserved baryon density can be small compared to the total energy density budget. Hence, defining the fluid velocity through the particle current, as in the Eckhart frame, is less useful~\citep{Denicol_2012}.
\subsection{MHD and force-free electrodynamics}
The ideal-MHD condition is~\citep{MT-1982}
\begin{align}
    F^\nu{}_\mu U^\mu
    =
    0
    \,,
\end{align}
which states that the electric field vanishes in the local fluid rest frame.
In the non-relativistic limit, this reduces to
\begin{align}
    \boldsymbol{E}
    +
    \boldsymbol{v}
    \times
    \boldsymbol{B}
    =
    0
    \,.
\end{align}
This condition does not imply that the Lorentz force on the fluid vanishes. To see this, consider the electromagnetic current
\begin{align}
    J^\mu
    &=
    \sum_s q_s N_s^\mu
    =
    \rho_{\rm q} U^\mu
    +
    j^\mu
    \,,
\end{align}
where
\begin{align}
    \rho_{\rm q}
    =
    \sum_s q_s n_s
    \,,
    \quad
    j^\mu
    =
    \sum_s q_s V_s^\mu
    \,,
\end{align}
and $U_\mu j^\mu=0$.
The spatial current can be further decomposed into components parallel and
perpendicular to the magnetic field,
\begin{align}
    j^\mu
    =
    j_{\parallel} \hat{b}^\mu
    +
    j_{\perp}^\mu
    \,,
\end{align}
For an ideal-MHD electromagnetic field,
\begin{align}
    F^\nu{}_\mu U^\mu
    =
    0
    \,,
    \qquad
    F^\nu{}_\mu \hat{b}^\mu
    =
    0
    \,.
\end{align}
It follows that
\begin{align}
    F^\nu{}_\mu J^\mu
    =
    \rho_{\rm q} F^\nu{}_\mu U^\mu
    +
    j_{\parallel} F^\nu{}_\mu \hat{b}^\mu
    +
    F^\nu{}_\mu j_{\perp}^\mu
    =
    F^\nu{}_\mu j_{\perp}^\mu
    \,.
\end{align}
Therefore, neither the charge-density contribution nor the field-aligned
current contributes to the Lorentz force in ideal MHD. The Lorentz force is
instead determined by the perpendicular current.

The force-free condition, by contrast, is
\begin{align}
    F^\nu{}_\mu J^\mu
    =
    0
    \,.
\end{align}
This distinction is particularly important in the KMHD expansion.
At leading order, each species satisfies $F^\nu{}_\mu N_s^\mu =0$
and therefore $F^\nu{}_\mu J^\mu = \sum_s q_s F^\nu{}_\mu N_s^\mu = 0$.
However, this is only a leading-order statement and does not imply that the
plasma is force free at every order in the KMHD expansion.
At subleading order, $F^\nu{}_\mu J^\mu$ is generally nonzero and contributes
to the plasma dynamics; see Eq.~\eqref{eq:all-species-equations-3+1}.